\documentclass[apj]{emulateapj}
\usepackage{amsmath}
\usepackage{amssymb}
\usepackage{epsfig}
\usepackage{apjfonts}

\newcommand{\cms}{\,{\rm cm$^{-2}$}\,}
\newcommand{\cmc}{\,{\rm cm$^{-3}$}\,}
\newcommand{\kms}{\,{\rm km\,s$^{-1}$}\,} 
\newcommand{\kmsmpc}{\,{\rm km\,s$^{-1}$\,Mpc$^{-1}$}\,}

\newcommand{\etal}{{ et~al.~}}

\newcommand{\sbunit}{\,{\rm counts\,cm$^{-2}$\,pixel$^{-1}$}\,}
\newcommand{\ergs}{\,{\rm ergs\,s$^{-1}$}\,}
\newcommand{\ergscm}{\,{\rm ergs\,s$^{-1}$\,cm$^{-2}$}\,}
\newcommand{\ergcmc}{\,{\rm ergs\,cm$^{-3}$}\,}

\newcommand{\Ms}{M_\odot}

\newcommand{\Zs}{Z_\odot}

\shorttitle{{\it Chandra} Observations of NGC~4782/4783}
\begin{document}
\title{X-ray Constraints on Galaxy-Gas-Jet Interactions in the
  Dumbbell Galaxies NGC~4782 and NGC~4783 in the LGG~316 Galaxy Group}

\author{M. E. Machacek, R. P. Kraft, C. Jones, W. R. Forman}
\affil{Harvard-Smithsonian Center for Astrophysics \\ 
       60 Garden Street, Cambridge, MA 02138 USA}
\author{M. J. Hardcastle}
\affil{School of Physics, Astronomy and Mathematics, 
  University of Hertfordshire \\
   College Lane, Hatfield, AL10 9AB, UK}

\email{mmachacek@cfa.harvard.edu}

\begin{abstract}
We present results from a $49.3$\,ks {\it Chandra} X-ray observation
of the strongly interacting dumbbell galaxies NGC~4782(3C278) and
NGC~4783 that constrain the kinematics of the interaction and models 
for bending the radio jets associated with NGC~4782. The galaxies are 
embedded in an approximately spherical distribution 
 of group gas, centered on NGC~4782, that, away from the interaction region,  
is nearly isothermal with mean temperature $kT \sim 1.4 \pm
0.4$\,keV. The X-ray morphology suggests that NGC~4783 is infalling 
into a single, massive galaxy group (LGG~316) with NGC~4782 nearly at 
rest at the center of the group potential. 
NGC~4783 shows a sharp X-ray surface brightness edge (cold front) 
to the east and  a $\sim 15$\,kpc ram-pressure-stripped tail to the west. 
Analysis of this cold front indicates NGC~4783 is 
moving east with a total velocity $870^{+270}_{-400}$\kms
(Mach  $1.4^{+0.5}_{-0.7}$) at an inclination angle
$46^\circ (>33^\circ)$   
towards us with respect to the 
plane of the sky. A $\sim 45$\,Myr
old X-ray cavity, with enthalpy of 
$4.4 \times 10^{57}$\,ergs, coincides with the eastern radio lobe of 3C278.
X-ray knots are found on both the radio jet and counter-jet,
coincident with peaks in the radio emission.
If we assume a light, mildly relativistic jet in 3C278, then ram pressure 
velocities of $100-200$\kms impacting the eastern jet and $\sim
170$\kms acting on the western radio lobe are sufficient to produce 
their observed bending. These velocities may be caused by bulk motions 
of the surrounding gas induced by the high velocity interaction between the 
galaxies, by the motion of the host galaxy NGC~4782 relative to the
IGM, or a combination of these processes. 

\end{abstract}



\section{INTRODUCTION}
\label{sec:introduction}

Deep, high angular resolution X-ray observations with {\it Chandra} and 
XMM-Newton have shown that the 
signatures of gas-dynamical interactions, i.e. shocks, cold
fronts, wakes or tails, provide
powerful new tools to constrain the three dimensional motion of the 
galaxies with respect to the surrounding IGM (Merrifield 1998;
Vikhlinin \etal 2001), as well as revealing the 
thermodynamic state of gas in and near the interacting system.   
In particular, a leading cold front - trailing wake morphology 
of a galaxy moving through the surrounding IGM fixes the direction 
of motion of the galaxy in the plane of the sky. Analysis of cold 
fronts, first applied to merging subcluster cores by Vikhlinin \etal
(2001), determine the three-dimensional velocity of galaxies  
infalling into nearby groups and clusters (e.g. Machacek \etal 2005a,
Scharf \etal 2005 for NGC1404 in the Fornax Cluster; 
Machacek \etal 2006 for NGC~4552 in the Virgo Cluster;  
Sun \& Vikhlinin 2005 for UGC~6697 in the Abell 1367 cluster; 
Sun \etal 2005 for NGC~1265 in the Perseus Cluster). The morphology
and thermodynamic properties of X-ray gas in the galaxy wakes and tails  
have been used to identify high velocity interactions and  
constrain the three dimensional motion of the galaxy 
(Machacek \etal 2005b for NGC~6872 in the Pavo Group), as have 
the properties of shocks (for a recent example, see Rasmussen \etal
2006 for NGC~2276 in the NGC~2300 group). In interacting galaxies 
where gravitational tidal effects are also important, these
gas-dynamical methods provide important additional constraints on the 
dynamics of the system that complement those obtained from modeling
the tidal stellar distortions.

Dumbbell galaxy systems (Wirth \etal 1982) can represent high
velocity, deeply  inter-penetrating encounters of two 
roughly equal mass elliptical galaxies, with characteristic projected 
separations of $\sim 10$\,kpc (Prugniel \& Davoust, 1990) and mean relative
radial velocity differences of $\sim 650$\kms (Valentijn \&
Casertano, 1990).   
In addition, dumbbells frequently host radio sources in one or both of 
the galaxies (e.g. Harris \& Roberts 1960; Wirth \etal 1982; 
Parma \etal 1991; Gregorini \etal 1994). The majority of radio-active 
dumbbells are found near the cores of rich, often dynamically young, 
 clusters (see, e.g. Beers \etal 1992 for Abell~400; Reid \etal 1998 
for Abell~3528). A few also have been found in galaxy groups. Since
groups, like galaxy clusters,  harbor dense X-ray gas 
(e.g. Schwartz, Schwarz \& Tucker 1980; Mulchaey \& Zabludoff 1998; 
Osmond \& Ponman 2004), 
hydrodynamical interactions between the galaxies, radio jets, and
group IGM in these dumbbells can be significant. 

In this work we present the results from a {\it Chandra} 
observation of the radio-active interacting dumbbell system containing 
elliptical galaxies NGC~4782 and NGC~4783, located  
near the center of the galaxy group LGG~316 (Garcia 1993), 
with line-of-sight velocities $v_r = 4618 \pm 31$\kms  and  
$v_r = 3995 \pm 22$\kms, respectively (Martimbeau \& Huchra, 2006).
Optical photometry shows that the central isophotes in each of the 
dumbbell galaxies are off-centered and the stellar velocity profiles 
distorted, signaling the effects of intense tidal interactions
between the pair (Borne \etal 1988, Madejsky \etal 1991, 
Madejsky 1992, Combes \etal 1995). Using different numerical methods, 
Borne \etal (1988) and Madejsky \& Bien (1993) obtained orbital 
solutions for NGC~4782 and NGC~4783 by fitting the observed 
isophotal shapes and stellar velocity distributions to simulations 
of tidal forces acting on the galaxies during the encounter. 
However, these solutions were not unique and, although these authors 
 agreed that the data were consistent with a high relative velocity
interaction viewed shortly after pericenter passage of the galaxies, 
they disagreed on the details of the orbital solution for the pair.

NGC~4782 harbors the FRI radio source 3C278 (Harris \& Roberts 1960) 
with two moderately bent radio jets and lobes (Baum \etal 1988) and a 
$5$\,GHz flux density for its point-like nucleus of $82$\,mJy. In 
contrast, VLA measurements give an upper limit on the $5$\,GHz flux 
density from the nucleus of NGC~4783, its dumbbell companion, of 
$< 0.5$\,mJy  (Hardcastle \etal in preparation).
Early simulations of
the evolution of NGC~4782's  radio jets assumed the orbital 
solution of Borne \etal (1988), based on modeling the 
observed stellar tidal distortions of the dumbbell assuming 
two approximately equal mass field galaxies initially infalling from
infinity. Hot ISM gas distributions, associated with each of
the galaxies, were introduced as free parameters in these simulations
to provide ram pressure to account for the observed bending of the 
radio jets (Borne \& Colina 1993).  
X-ray observations allow a direct measurement of the density distribution 
of hot gas in and surrounding each galaxy, reducing the number of 
free parameters needed by simulations. X-ray
observations may also constrain the motions of the interacting
galaxies, allowing a more realistic choice of initial conditions 
for simulating the interaction. 
The presence of hot X-ray gas in
NGC~4782/NGC~4783 was first identified
using the Einstein Observatory (Roberts \etal 1991; Fabbiano \etal
1992). Later {\it ROSAT} HRI observations revealed a highly asymmetric 
hot gas distribution, 
with extended emission to the west of NGC~4783 and to the east of 
NGC~4782, and a bright bridge of X-ray 
emission connecting the two galaxy centers (Colina \& Borne 1995).
In this work we use {\it Chandra} observations of NGC~4782/NGC~4783 
to measure the X-ray gas properties both in the interacting galaxies and the 
surrounding group, in order to constrain the dynamics
of the system and the evolution of the radio features. 

This paper is organized as follows.  In \S\ref{sec:obs}
we discuss the observation and our data analysis procedures.
In \S\ref{sec:galint} we examine the effects of the high velocity
encounter of NGC~4783 with NGC~4782. We first measure the thermodynamic 
properties of hot gas in the inner regions of the system 
(\S\ref{sec:galspec}) and then  characterize the mass,
density and temperature structure of the larger scale group
(\S\ref{sec:group}). In \S\ref{sec:cavities} we discuss the  
X-ray features associated with the radio jets.
In \S\ref{sec:velcon} we use these data to constrain the kinematics of
the galaxy interactions with each other and with the IGM, and 
present a physical picture for NGC~4782/4783 dumbbell interaction 
suggested by our analysis in \S\ref{sec:discuss}.
We summarize our primary findings in \S\ref{sec:conclude}.
Throughout this paper we adopt a flat, 
dark energy dominated cosmology ($\Omega_m = 0.3$, $\Lambda=0.7$) with 
Hubble parameter $H_0 = 70$\kmsmpc. 
Taking the redshift of NGC~4782 ($z=0.0154$, 
Martimbeau \& Huchra 2006) as representative of the 
group, the luminosity distance to the system is $66.7$\,Mpc and $1''$
corresponds to a distance scale of $0.314$\,kpc.
Unless otherwise indicated, all errors correspond to $90\%$ 
confidence levels and coordinates are J2000.0.
\section{OBSERVATIONS AND DATA REDUCTION}
\label{sec:obs} 

\subsection{Chandra Observations}
\label{sec:chandraobs}

 {\it Chandra} observed the dumbbell galaxies NGC~4782 and NGC~4783 
(obsid 3220) on 2002 June 16 for $49.3$\,ks using the Advanced CCD 
Imaging Spectrometer (ACIS, Garmire \etal 1992) in VFaint mode 
with the back-illuminated chip S3 at the aimpoint. The data were 
reprocessed using standard X-ray analysis packages (CIAO 3.2,  
FTOOLS). Events were filtered to remove bad grades 
($1$, $5$, $7$), and hot pixels and columns. Additionally, 
we used VFaint mode filtering to obtain a factor  
$\sim 2-3$ better rejection of particle backgrounds at energies 
below $1$\,keV. The data 
were reprocessed using updated gain tables and corrected for the slow 
secular drift of the average pulse-height amplitude (PHA)  values for 
photons of fixed energy 
(tgain\footnote{see Vikhlinin \etal in
 http://cxc.harvard.edu/contrib/alexey/tgain/tgain.html}) 
and for the buildup of contaminants on the optical filter 
(Plucinsky \etal 2003). Periods of anomalously high
background (flares), as well
as those with anomalously low count rates,  
were removed using the script lc\_clean on the S3 data 
in the $2.5-7$\,keV energy band after masking out bright
sources. This resulted in a useful exposure time of
$46,667$\,s. 

We created backgrounds for imaging and  
spectral analyses of the group IGM from the $450$\,ks 
period D source free dataset
aciss\_D\_7\_bg\_evt\_010205.fits appropriate for the
date of observation and instrument 
configuration\footnote{ see http://cxc.harvard.edu/contrib/maxim/acisbg}. 
We checked  the normalization of the source free background by comparing 
count rates in the source and background files in the $9.0-11.5$\,keV
energy bands, where particle background dominates. We 
found that these count rates agreed to better than $1\%$, 
such that no additional background renormalization was required. 
 
We used a multiscale wavelet decomposition algorithm with a 
$5\sigma$ detection threshold to identify X-ray point sources in 
the $8'.4 \times 8'.4$ field of view of the S3 chip in four 
energy bands: $0.3-1$\,keV (soft), $1-2$\,keV (medium), 
$2-8$\,keV (hard), and $0.3-8$\,keV (broad). We detected $59$ 
X-ray sources on the S3 chip, in addition to the nucleus of NGC~4782, 
with $29$ of them lying within a $2'$ circle centered on NGC~4782. 
Three of these sources (two in the western jet and one in the eastern 
jet) are X-ray knots coincident with peaks 
in the radio emission and are discussed in \S\ref{sec:jetemiss}. The 
bright point source at ($12^h54^m35.41^s$,$-12^\circ34'5.38''$),
located just north of NGC~4782's nucleus   
and containing $51 \pm 7$ counts ($1\sigma$ uncertainties) in the 
$0.3-8$\,keV energy band, is likely a ULX in the core of NGC~4782. 
For power law spectral photon indices $\Gamma \sim 1.6 - 2$ and 
Galactic absorption, we find a $0.3-8$\,keV luminosity of 
$(3 \pm 1) \times 10^{39}$\,ergs for this ULX.  
All point sources, except for the nucleus of NGC~4782, were removed
from the surface brightness and spectral analyses. 

\subsection{VLA Data}
\label{sec:vlaobs}

To complement the {\it Chandra} data, we obtained previously
unpublished deep observations from the archive of the NRAO Very Large
Array (VLA), made at effective frequencies of $1.5$ and $4.9$\,GHz. Details
of these observations are given in Table \ref{vla-data}.

The VLA data were reduced in the standard manner within {\sc aips}.
Data from each individual observation were calibrated and
self-calibrated; images made from long-baseline data were then used to
cross-calibrate the short-baseline data and the datasets were
concatenated to make a single dataset from which maps at various
resolutions could be derived. In this paper we use the $1.5$-GHz images
to show the large-scale structure of the source, while the higher
resolution available at $4.9$\,GHz is used to show details of the jet.

The VLA data, together with our new Giant Meter-wave Radio Telescope
(GMRT) observations of 3C\,278, will be discussed in more detail in a
future paper.

\section{The High Velocity Encounter of NGC~4782 and NGC~4783}
\label{sec:galint}

\begin{figure}[t]
\includegraphics[height=2.09in,width=3in]{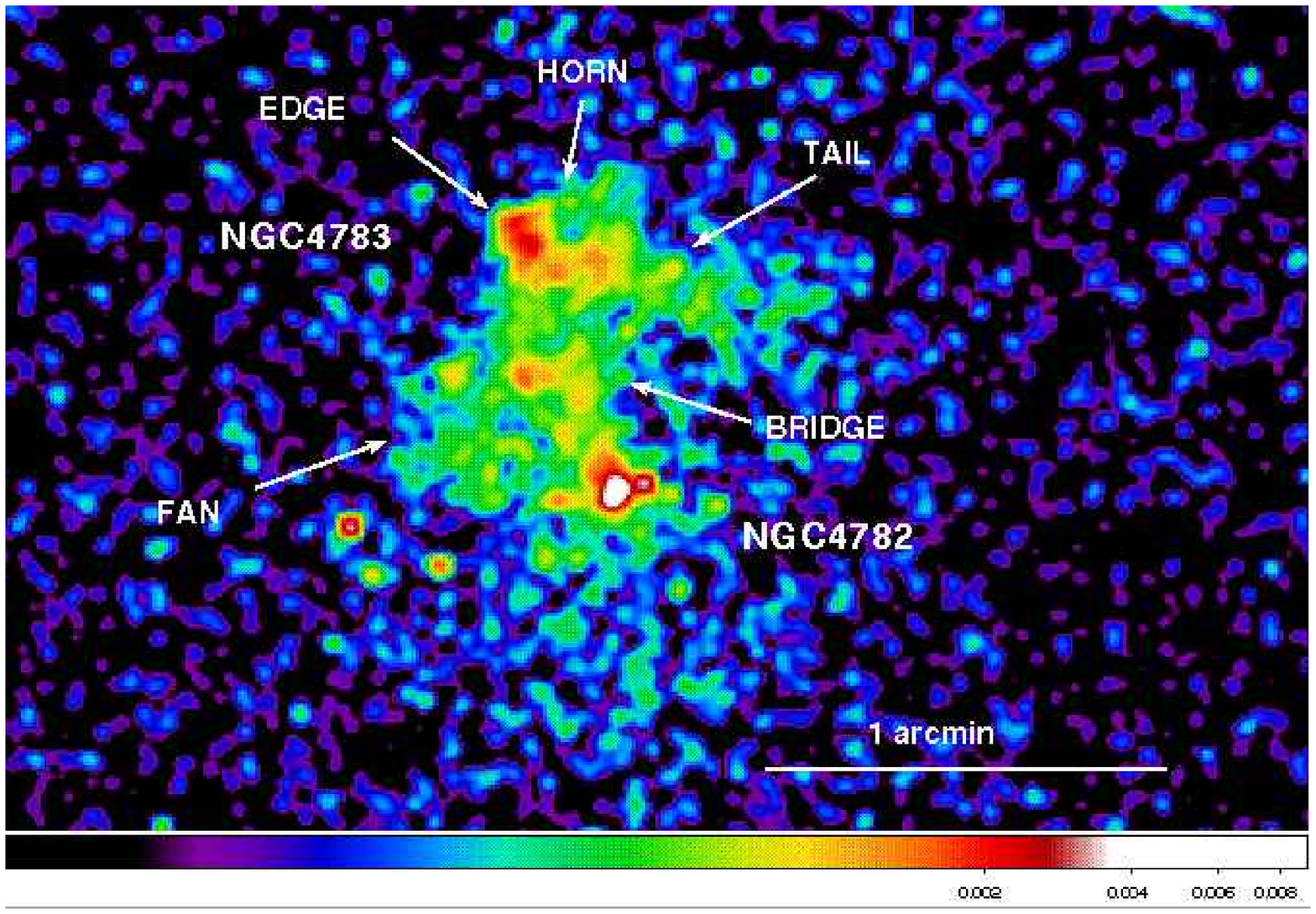}
\includegraphics[height=2.26in,width=3in]{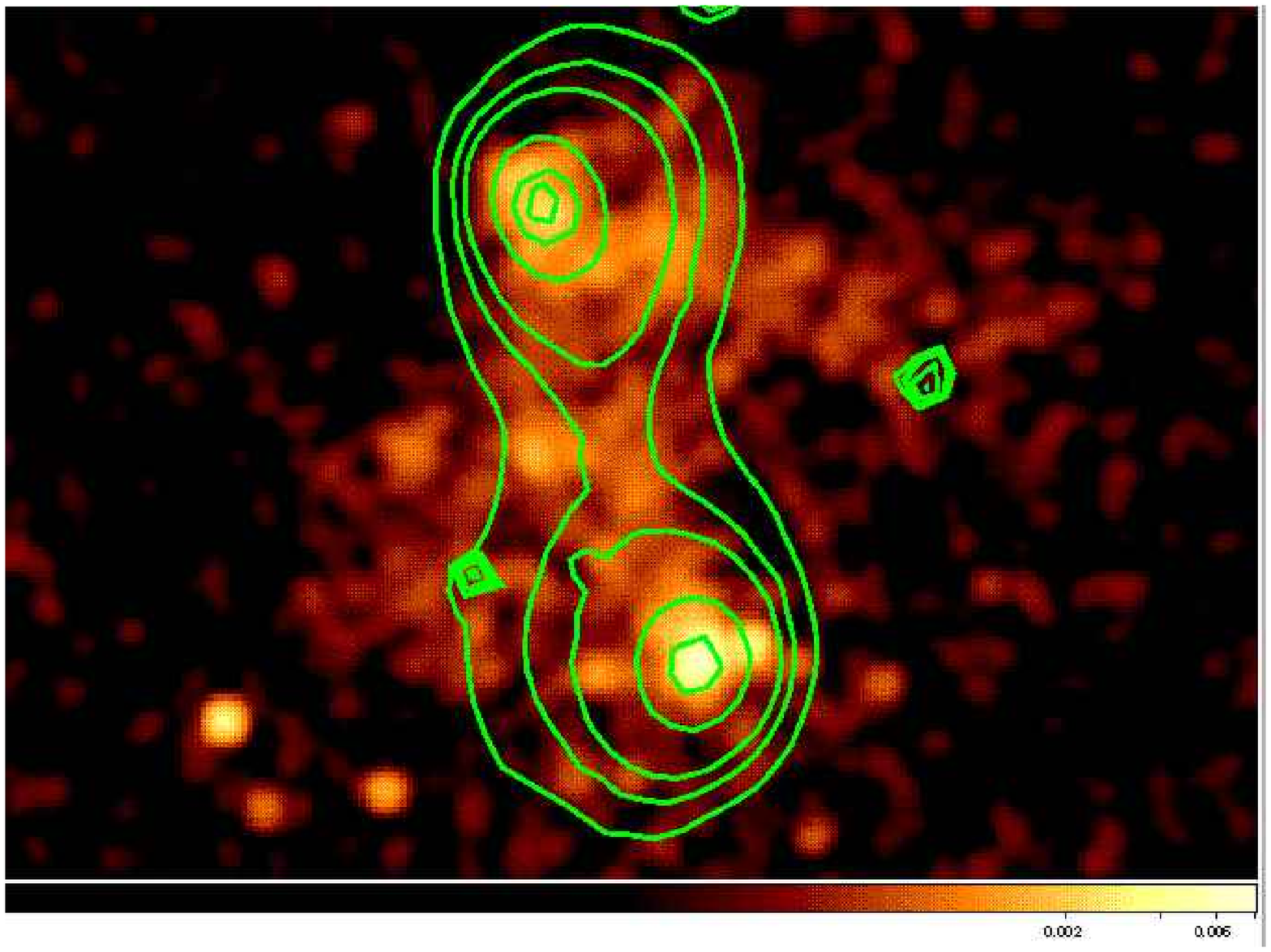}
\caption{ ({\it top}) $0.3-2$\,keV\,{\it Chandra} image of the
dumbbell galaxies NGC~4782 and NGC~4783. The image has been background
subtracted, exposure corrected and smoothed with a $1''$ Gaussian
kernel.  $1\,{\rm pixel} = 0''.49 \times 0''.49$.
({\it bottom})  $0.3-2$\,keV\,{\it Chandra} image (as in the left panel)  
with  optical contours superposed from a 
WFPC2 (FR680N) Hubble Space Telescope image. Contour levels correspond
to flux densities of ($3$, $4.5$, $6$, $12$, $24$, $48$) 
$\times 10^{-21}$\ergscm \,$\stackrel{\circ}{A^{-1}}\,{\rm pixel}^{-1}$, respectively.
}
\label{fig:chandra}
\end{figure}

In Figure \ref{fig:chandra} we show the 
$0.3-2$\,keV {\it Chandra} X-ray image of the interacting galaxies 
NGC~4782 and NGC~4783. The X-ray image has
been background subtracted, corrected for telescope vignetting 
and detector response, and then smoothed with a $1''$ Gaussian kernel. 
In the bottom panel of Figure \ref{fig:chandra}, we show isophotes from a 
WFPC2 (FR680N) Hubble Space Telescope image of NGC~4782 and NGC~4783 
superposed on the $0.3-2$\,keV {\it Chandra} image.

We see that the peaks of 
the X-ray emission coincide with the off-centered
peaks in the optical light at the locations of the galaxies' nuclei.
Off-centered stellar isophotes are the defining signature of 
strong tidal interactions between elliptical   
galaxies, when the distance of approach is smaller than the
characteristic size of the system. The off-centering 
results from  the tidal deceleration 
of the outer stellar envelopes of each galaxy towards its partner, 
while stars near the galaxy centers have had time, due to their 
short ($\sim 10^7$\,yr) dynamical time scales, to average out the 
tidal effects and return to equilibrium (Madejsky \etal 1991, 
Combes \etal 1995). More asymmetric extended
features, such as tidal tails, are not yet seen, since the stars 
which became unbound during the pericenter passage of the galaxies, 
have not had time to escape the parent galaxies (Combes \etal 1995). 

A key feature that distinguishes gravitational interactions (tides) 
from gas-dynamical processes (ram-pressure or turbulent-viscous 
stripping) is that gravitational interactions act on both stars and
gas, while gas-dynamical processes are ineffective at distorting 
the stellar distributions or velocities. 
From the bottom panel of Figure \ref{fig:chandra}, we see  
an enhanced `bridge' of X-ray emission, along the line 
(oriented $19^\circ$ east of north) connecting the galaxy centers. 
The `bridge' is closely correlated with the stellar distribution,  
as shown by the HST isophotes (overlaid contour lines),  
and coincides with the direction of the north-south tidal 
distortions in the stellar isophotes. This suggests that the 
formation of the X-ray `bridge' is due in part to the same tidal
interactions that disturbed the stars. Emission in the `bridge' 
region is also likely partly due to projection effects from the 
superposition of the two galaxy ISMs caused by the closeness of 
the encounter. 

The X-ray surface brightness distribution has extended 
features that do not follow the stellar distribution, suggesting 
that hydrodynamical processes are important. In the top panel of 
Figure \ref{fig:chandra}, NGC~4783 shows a sharp
surface brightness discontinuity (edge) on the eastern side of the 
galaxy, a $\sim 4-5$\,kpc long streamer or `horn' of gas 
swept back from the edge to the northwest, 
and a tail of enhanced X-ray emission, extending 
$\sim 15$\,kpc to the west, opposite NGC~4783's `leading' edge. These
features are similar to those predicted in numerical simulations of
gas-dynamical stripping of elliptical galaxies due to their motion 
through the ambient IGM (Stevens \etal 1999; Toniazzo \& Schindler
2001; Acreman \etal 2003; Heinz \etal 2003), and to features seen in 
other elliptical galaxies undergoing ram-pressure and/or turbulent-viscous
stripping. The leading edge - trailing tail morphology determines 
that NGC~4783's motion in the 
plane of the sky is towards the east. This is in 
qualitative agreement with the relative galaxy motions from 
the simulations of Borne \etal (1988) and Madejsky \etal (1993), 
where NGC~4783 passes to the west of NGC~4782 at pericenter. 
The filamentary `horn' extending from the sharp edge of 
NGC~4783 is likely gas being stripped at the galaxy-IGM interface. 
The twisted morphology 
of the bright inner regions of NGC~4783's tail is reminiscent of 
von Karman vortices seen in hydrodynamic fluid flows  and produced in
 recent simulations of ram pressure stripping of 
disk galaxies (Roediger \etal 2006).

An asymmetric  `fan' of X-ray emission extends to the 
east-northeast of NGC~4782, 
overlapping in projection the bend in the eastern radio jet 
(see the bottom panel of Fig. \ref{fig:cavity}), while less X-ray
emission is observed to the southwest. However, identifying the 
dominant processes responsible for the origin of the fan is complex. 
Madejsky (1992) found a sharp increase in the stellar velocity 
dispersion $\sim 15-20''$ from NGC~4782, in the base of the fan, 
indicating the strong influence of tidal forces on matter (gas and
stars) in the region. The compression wave produced by the 
high speed  passage of the companion galaxy NGC4783 from the 
west to the northeast of NGC~4782 would displace the ISM outside 
the core of NGC~4782 to the east relative to NGC~4782's stellar 
distribution, as is seen. Similar bulk
 motions, on larger scales, have been shown to occur in  
 ICM gas in merging clusters (Ascasibar \& Markevitch 2006).
Any small velocity of NGC~4782 relative to the
group potential and ambient IGM, induced by the tidal interaction of
the galaxy pair, also would contribute to the hydrodynamical displacement 
of the gas. Finally, as we show in \S\ref{sec:jetemiss}, some of 
the X-ray emission might be associated with the eastern radio jet. 

To probe the  
evolution of NGC~4782, NGC~4783 and the group IGM, we next characterize 
the thermodynamic properties of the hot gas in the galaxies 
and tail, bridge and fan (\S\ref{sec:galspec}), as well as in the 
surrounding group IGM (\S\ref{sec:group}).  

\subsection{Galaxy Gas Temperatures and Densities}
\label{sec:galspec}

\begin{figure}[t]
\includegraphics[height=2.68in,width=3in]{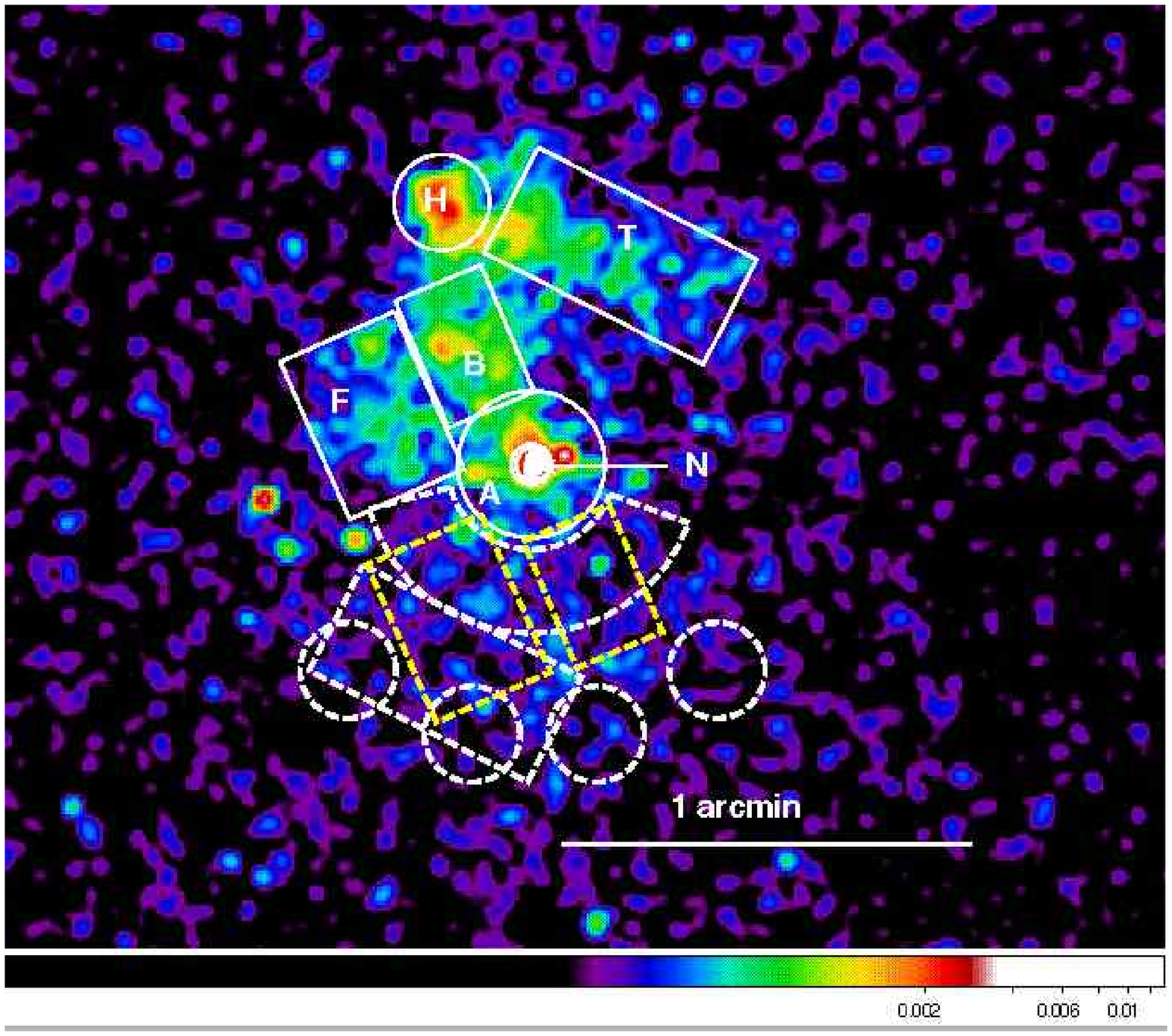}
\includegraphics[height=2.6in,width=2.6in]{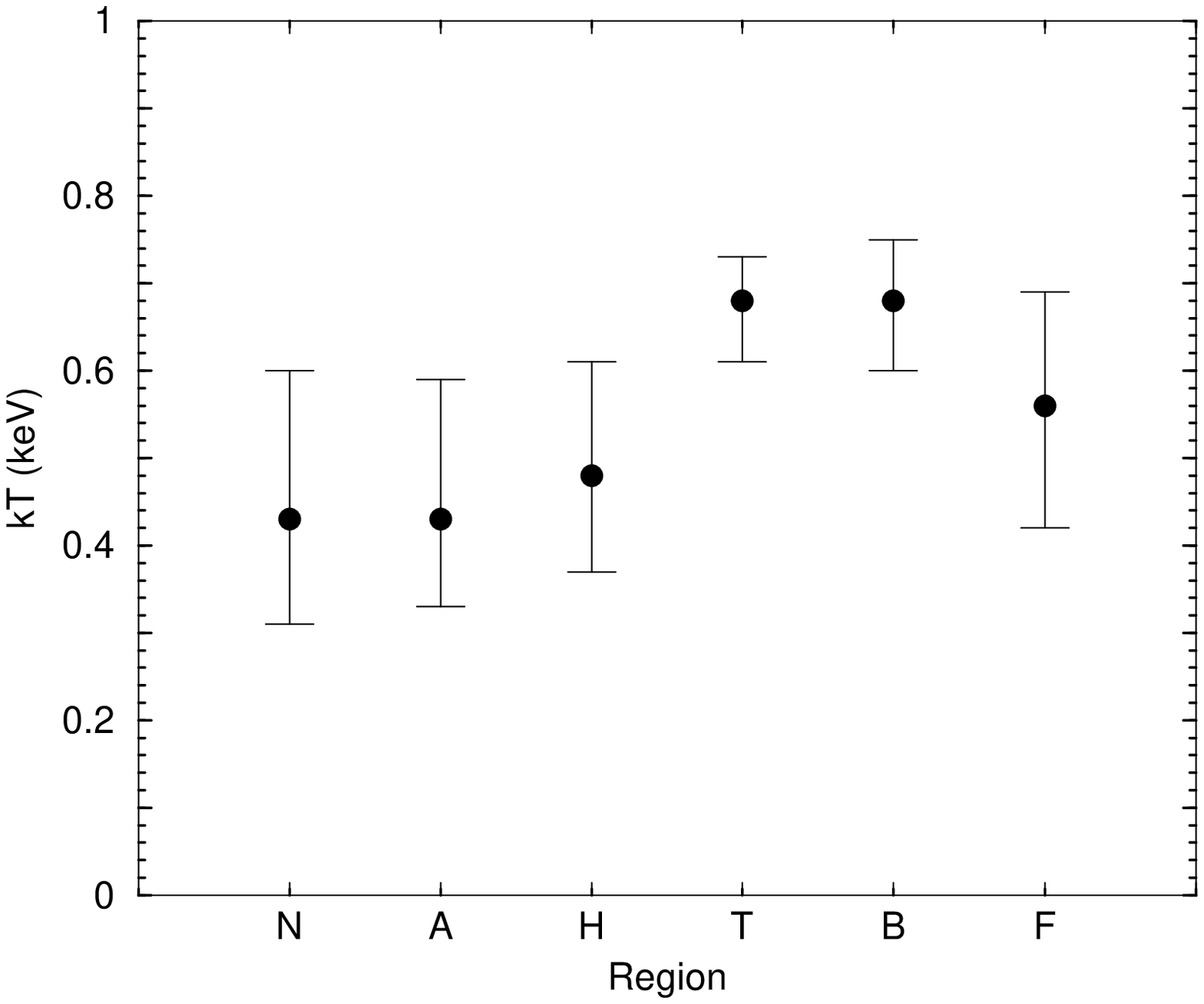}
\caption{ ({\it top}) $0.3-2$\,keV background subtracted, exposure
  corrected and Gaussian smoothed {\it Chandra} image of the
dumbbell galaxies NGC~4782 and NGC~4783 with spectral regions for 
N4783's head (H) and tail (T), the interconnecting bridge (B), 
eastern fan (F), and NGC~4782's nucleus (N) and surrounding annulus (A)
superposed. Local background regions (dashed lines) are also shown for
  region H (circles), A (annular sector), T(white box), B(right yellow
  box), and F(left yellow box).
({\it bottom}) Gas temperatures in the elliptical galaxies 
NGC~4782 and NGC~4783 for the spectral regions
  shown to the left and spectral models listed in 
Table \protect\ref{tab:galspectra.brems.2}. 
}
\label{fig:galspecv2}
\end{figure}

The spectral extraction regions used to determine the properties of 
hot gas in the elliptical galaxies NGC~4782 and NGC~4783 and in the 
nearby extended X-ray features are listed in Table
\ref{tab:specregsv2} and are shown in the top panel of 
Figure \ref{fig:galspecv2}.
We use local background regions with the same geometries and at the same 
mean radial distances as the regions of interest to subtract the 
contribution of the surrounding group IGM, as well as the particle,
soft Galactic, and cosmic X-ray backgrounds, from the galaxy spectra 
for regions H, T, B, and F. For the annular region A in NGC~4782, we use an  
annular sector abutting region A to the south. These local background 
regions are also listed in Table \ref{tab:specregsv2} and delineated with
dashed lines in the bottom panel of Figure \ref{fig:galspecv2}. 
For the background of the nuclear region N of NGC~4782, we used 
a small annulus (region A in Fig. \ref{fig:galspecv2}).
Counts were grouped for spectral fitting with a minimum $20$ counts 
per bin.  All spectra were then fit using XSPEC 11.3 with instrumental 
response matrices produced using standard CIAO tools. We restricted
our spectral fits to the $0.3-8$\,keV energy range for regions 
H, T, and F, $0.3-2$\,keV for regions B and A, and $0.3-5$\,keV for 
the NGC~4782 nucleus (region N). 
We discuss our spectral analysis for all regions outside the nucleus below, 
and for the nuclear region in \S\ref{sec:nucagn}.

We used single temperature APEC thermal plasma 
models (Smith \etal 2001) with solar abundance tables from 
Anders \& Grevasse (1989) to model the hot ISM. 
Each model component was corrected for absorption 
using Wisconsin photo-electric cross-sections (Morrison \& McCammon 1983). 
We found no evidence for increased absorption over  
the Galactic value, and so fixed the 
hydrogen absorbing column at $3.58 \times 10^{20}$\cms 
(Dickey \& Lockman 1990) for all model 
components. Since abundances in nearby elliptical galaxies are known
to vary, we explore models with fixed metal abundance between 
$0.3 - 1.2\Zs$, where the lower bound is taken from the $90\%$ CL lower
limit on the metallicity of the surrounding IGM (see \S\ref{sec:group}
and Table \ref{tab:igmfitswO}) and the upper bound corresponds to the highest
metallicities observed in comparable,  moderate-X-ray-luminous 
elliptical galaxies in the sample of Humphrey \& Buote (2006). 

Our regions contain considerable stellar mass, such that
the contribution from LMXBs may be significant. We correct for the
presence of LMXBs following 
Gilfanov (2004), who showed that the X-ray luminosity
function for LMXBs in elliptical galaxies has a universal shape with 
its normalization proportional to the galaxy's stellar mass. We 
determine the contribution of LMXBs to the $0.5-8$\,keV X-ray
luminosity 
from the stellar mass in each spectral
region, determined from K-band photometry. 
The contribution of LMXBs to each X-ray spectrum is modeled by 
adding a $5$\,keV 
bremsstrahlung component to the spectral models (Kraft \etal 2001) 
with its normalization fixed by the expected LMXB unabsorbed X-ray flux. 
We checked that our results were unchanged if we instead chose to model 
the LMXB contribution to the spectra with a ($\Gamma = 1.6$ ) power
law.

The results of our spectral fits and the intrinsic X-ray luminosities  
for each region are given in 
Tables \ref{tab:galspectra.brems.2} and \ref{tab:lums}, respectively. 
Only in two regions (H and A) does the $0.5-2$\,keV emission from
LMXB's contribute more than $20\%$ of the total emission.
We find that,  once the contribution of LMXBs has been taken into
account, single temperature APEC models for hot gas provide excellent fits 
to the remaining X-ray emission in all regions outside the nuclear 
region (N). The best fit temperatures for 
these regions do not change when the abundances are varied between 
$0.3-1.2\Zs$. Similarly, the derived $0.5-2$ keV and total $0.5-10$\,keV X-ray
luminosities vary by $\lesssim 4\%$, and the  $2-10$\,keV luminosity by 
$\lesssim 17\%$, over the abundance range of interest. Thus, for
simplicity, we present only the $0.5\Zs$ metallicity model results 
in Tables \ref{tab:galspectra.brems.2} and \ref{tab:lums}. 
We plot the best fit gas 
temperatures as a function of region label in the bottom panel of 
Figure \ref{fig:galspecv2}. 
The gas temperature in the central regions of the
galaxies ($0.48^{+0.13}_{-0.11}$\,keV for region H in NGC~4783 and 
$0.43^{+0.16}_{-0.10}$\,keV for the annular region A in NGC~4782) are
typical of those found in the cores of elliptical galaxies. 
Gas temperatures in the extended features, 
$0.68^{+0.05}_{-0.07}$\,keV for NGC~4783's tail
region T, $0.68^{+0.07}_{-0.08}$\,keV for the bridge B, and 
$0.56^{+0.13}_{-0.14}$ for the fan F  
are typical of galaxy gas that has been stripped by hydrodynamical or 
tidal processes and significantly cooler than the temperature measured 
for the group IGM (see \S\ref{sec:grouphydro}). 

We use the best fit spectral models  
to estimate the mean density and gas mass in 
each region  
(see, e.g. Eq. 3 in Machacek \etal 2006), and to calculate
the thermal gas pressures. For the tail 
region (T), bridge (B) and fan (F), we assume gas  
fills a cylinder of rotation about the major axis with (projected) length
 $l_p$ and diameter given in Table \ref{tab:specregsv2}. We assume 
spherical volumes for the central regions of the galaxies. 
Our results for the electron densities,
thermal pressures, cooling times, and total gas masses, assuming 
uniform filling and zero inclination angle with respect to the 
plane of the sky, 
are listed in Table \ref{tab:galdens.brems.2}, 
where the uncertainties reflect 
both the $90\%$ CL uncertainties in the temperature from the spectral 
fits and the uncertainties in the assumed metal abundance. The upper
and lower uncertainties on the density, pressure and gas mass  
are derived from the low ($0.3\Zs$) and high ($1.2\Zs$) abundance 
models, respectively. The mean electron density in each region 
also depends upon the relative geometry of the collision through 
the inclination angle $\zeta_i$ 
of each region with respect to the plane of the 
sky. The physical length (and thus volume) of each cylindrical 
region (T, B, and F) scales as ${\rm cos}^{-1}(\zeta_i)$, 
such that the gas density and pressure decrease with increasing angle of 
inclination, scaling as ${\rm cos}^{1/2}(\zeta_i)$, while  
the total gas mass increases, scaling as  
${\rm cos}^{-1/2}(\zeta_i)$.

For zero inclination angle and uniform filling, we find 
mean electron densities  
$n_e \sim 0.01-0.03$\cmc, gas pressures 
$p \sim (1 - 5) \times 10^{-11}$\,ergs\,\cmc,
and cooling times $\sim 0.2 - 1.2$\,Gyr
(long compared to the timescale of the collision), 
for all regions outside the nuclear region N. Hot gas
masses in the central regions of the galaxies are 
$\sim 2.9^{+0.6}_{-1.0} \times 10^{7}\Ms$ in the central $2$\,kpc of 
NGC~4783 and 
$\sim 4.4^{+1.1}_{-1.5} \times 10^{7}\Ms$ in the central 
$3.4$\,kpc of NGC~4782, comparable to gas masses 
($3.6 \times 10^7\Ms$ for NGC~4783 and 
$8 \times 10^7\Ms$ for NGC~4782) found by direct integration of the 
$\beta$-model density distribution fits of Jones \etal (2007). 
If we compare these central gas masses to those within NGC~4472, a 
noninteracting elliptical
galaxy with similar velocity dispersion, we find
that the central gas masses in NGC~4472 (i.e. $6.8 \times 10^{7}\Ms$
and $1.43 \times 10^{8}\Ms$ within $r = 2$ and $3.4$\,kpc,
respectively, of NGC~4472's nucleus; Jones \etal 2007) are a factor 
$\sim 2 - 3$ larger  than the gas masses found in the central regions of the 
interacting galaxies NGC~4782 and NGC~4783, suggesting that gas may
have been depleted in the central regions of the dumbbell galaxies 
as a result of the interaction. 

For nonuniform filling ($\eta_i < 1$,
for each region $i$) the gas density and pressure are increased by a factor
$(\eta_i)^{-1/2}$ and total mass reduced by $(\eta_i)^{1/2}$. We
caution the reader that in regions, such as region F, where the
emission is clumpy and the filling factor likely $< 1$, the values in 
Table \ref{tab:galdens.brems.2} represent lower bounds on the density
and pressure and an upper bound on the derived X-ray gas 
mass in the region. 

\subsection{The AGN at the Heart of NGC~4782}
\label{sec:nucagn}

Our spectral fits to the nuclear region N of NGC~4782, that 
hosts the radio source 3C\,278, are listed 
in Table \ref{tab:nucspecfit}. Neither a single temperature APEC 
model nor a single power law model can describe the
spectrum. We obtained a good representation of the spectrum using an
absorbed power law plus thermal APEC model. 
Due to our low statistics ($193 \pm 14.5$ source counts), we 
fix the hydrogen absorption column at Galactic ($3.58 \times
10^{20}$\cms) and explore a range of photon indices ($\Gamma =
1.2 - 2.2$) and gas metallicities ($0.3-1.5\Zs$), typical for 
LLAGN (see, e.g. Evans \etal 2006 and references
therein), and gas in the centers of elliptical galaxies 
(Humphrey \& Buote 2006). Since the temperature of the gas is 
unchanged (at $90\%$ CL) when the power law photon index ($\Gamma$) and 
gas metal abundance are varied over these ranges, we list only a single case 
($\Gamma = 1.7$ and metal abundance $0.5\,Zs$) in 
Table \ref{tab:nucspecfit}.  We find a temperature 
$kT=0.4^{+0.2}_{-0.1}$\,keV, signaling the presence of cool X-ray
emitting gas within $0.7$\,kpc of NGC~4782's nucleus 
(See Fig. \ref{fig:galspecv2}). The best fit temperature for the 
thermal component is also unchanged when the bremsstrahlung component, 
accounting for LMXB's, is added to the model. 

In the nuclear region of NGC~4782, the total $0.5-10$\,keV 
luminosity of 
$(1.5 \pm 0.3) \times 10^{40}$\ergs 
is dominated by the power law component
with $(9 \pm 2) \times 10^{39}$\ergs,
$4.2^{+0.8}_{-0.5} \times 10^{39}$\ergs
and $2.1 \times 10^{39}$\ergs 
attributed to the power law, thermal (gas), and bremsstrahlung (LMXB) 
components, respectively.  Since power law  X-ray emission  
may be produced by subparsec scale jets in the core of radio sources, 
as well as by accretion onto the central black hole (Evans \etal 2006), 
the observed power law X-ray luminosity 
( $(9 \pm 2) \times 10^{39}$\ergs)  
is an upper bound on the X-ray luminosity of the central engine.

We estimate the mass ($M_{BH}$) of the central black hole in NGC~4782 
to be $10^9\Ms$ using the central stellar velocity dispersion 
of $310$\kms (Madejsky \etal 1991; Madejsky 1992) 
in the $M_{BH} -\sigma_c$ relation (Ferrarese \& Merritt 2000; 
Gebhardt \etal 2000). The  Eddington luminosity for such 
a massive black hole is $L_{\rm Edd} \sim 1.3 \times 10^{47}$\ergs,
 such that  $L_{\rm X}/L_{\rm Edd} < 10^{-7}$.
Even if the bolometric luminosity is $3-10$ times higher than the 
observed $0.5-10$\,keV X-ray luminosity, as is the
case for accretion-dominated AGN (Elvis \etal 1994), the central AGN
in NGC~4782 is radiating at very low luminosity for its mass.

Variation of the metal abundance does, however, affect the APEC
component normalization, which decreases as the model metallicity is
increased, and thus the derived gas density, thermal pressure and 
gas cooling times. These quantities remain insensitive to the 
variation of the power law photon index. For all quantities below, 
the quoted uncertainites reflect both the uncertainties due to the 
variation of gas metal abundance and LLAGN photon index in the
spectral model and the $90\%$ CL uncertainties in the  
gas temperature in the model fit.  
We find a  mean electron density 
$n_e \sim 0.09 \pm 0.04$\cmc and 
thermal pressure 
$p \sim (1.2^{+1.0}_{-0.7}) \times 10^{-10}$\,ergs\,\cmc.
The cooling time
($\sim 80^{+50}_{-40}$\,Myr) in region N is short, 
suggesting  that gas accretion onto the supermassive black hole can  
trigger episodes of AGN jet activity.

Given the density and temperature of the gas in the nuclear region, 
we can estimate the power available 
from Bondi accretion ($\dot{M}_{\rm BH}c^2$) onto the central black hole, 
where $c$ is the speed of light. The Bondi accretion rate is 
\begin{multline}
 \dot{M}_{\rm BH} = 7 \times 10^{23}\\
\times (M_{\rm BH}/10^9\Ms)^2(n_e/0.17{\rm cm}^{-3})(kT/0.8\,{\rm keV})^{-3/2}\,{\rm g\,s}^{-1} 
\label{eq:bondi}
\end{multline}
 (di Matteo \etal 2003).   
Using the density 
($0.09 \pm 0.04$\cmc) and temperature
($0.4^{+0.2}_{-0.1}$\,keV) from region N, we find a Bondi accretion rate of 
$\dot{M}_{\rm BH} \gtrsim 0.020^{+0.025}_{-0.015}\Ms\,{\rm yr}^{-1}$. 
This is a lower bound on the Bondi accretion rate, because 
the Bondi accretion radius ($r_a = GM_{\rm BH}/c_s^2$) is much 
smaller than the $0.7$\,kpc radius of region N, such that the 
mean density from our spectral fit to region N likely 
underestimates the gas density at the 
accretion radius.  Assuming the lower bound on the Bondi accretion
rate ($0.005\Ms\,{\rm yr}^{-1}$) from the lower limit 
 on the gas density (metal abundance $1.5\Zs$) and upper $90\%$ confidence 
limit on the temperature, the minimum total power 
available from Bondi accretion ($\dot{M}_{\rm BH}c^2$) is then 
 $\gtrsim 3 \times 10^{44}$\ergs. 
Assuming the standard radiative efficiency of $0.1$, 
we would expect an accretion luminosity 
onto the central black hole in NGC~4782 greater than $3 \times 10^{43}$\ergs. 
The observed X-ray luminosity is 
more than $3000$ times smaller.
The inferred radiative efficiency for   
the central AGN, $L_{\rm X}/\dot{M}_{\rm BH}c^2 < 3 \times 10^{-5}$,  
is then an upper bound to the true radiative efficiency, both because the 
Bondi accretion rate may be underestimated
and because the observed nuclear X-ray luminosity for FRI sources 
is likely contaminated by emission from a subparsec scale jet (Evans
\etal 2006).   

\subsection{Characterization of the Group Environment}
\label{sec:group}

NGC~4782 and NGC~4783 are not an isolated galaxy pair, but are part of
the LGG~316 galaxy group (de Souza \& Quintana 1990; Quintana \etal
1996). The derived properties of LGG316 are summarized in Table
\ref{tab:grpprop} and discussed below.

\subsubsection{Group Velocity Dispersion and Dynamical Mass}
\label{sec:grpdynmass}

We adopt a radial velocity of 
$4618 \pm 31$\kms for NGC~4782 and $3995 \pm 22$
for the northern companion NGC~4783 (Martimbeau \& Huchra 2006), 
in agreement with Borne \etal (1988), Madejsky \etal (1992), and 
Quintana \etal (1996). From $22$ galaxies with radial velocities 
between $3500$\kms and $5500$\kms in a $1^\circ$ square region 
centered on NGC~4782, we found a mean radial 
velocity of $4604$\kms and line-of-sight velocity dispersion of 
$447 \pm 70$\kms. Then using the mass-velocity dispersion ($M-\sigma$) 
relationship normalized by simulations  
($f_\sigma \sim 0.9$, Bryan \& Norman 1998), 
we estimate a dynamical mass within $r_{500}$ for the group of 
$\sim 7 \times 10^{13}\Ms$ with an uncertainty 
of $\sim 30\%$. These results are in excellent agreement with 
previous results (de Souza \& Quintana 1990; Quintana \etal 1996) 
when the galaxies are considered members of a single galaxy group (LGG~316) 
with NGC~4782 at its center.   
However, Quintana \etal (1996) also suggest that their data 
are consistent with a double-peaked galaxy distribution, in which 
NGC~4782 and NGC~4783 are the dominant members of two unequal mass 
subgroups undergoing a high velocity collision and 
possible merger. 

\subsubsection{Surface Brightness Profile, Temperature and Density}
\label{sec:grouphydro}

The characterization of the hydrodynamical interactions, acting to 
transform the galaxies during their encounter, requires 
a detailed understanding of the thermodynamic properties of the 
surrounding group IGM. Furthermore, X-ray measurements 
of the hot gas density and temperature distributions in the group IGM 
may reliably trace the underlying dark matter potential and 
reflect the dynamical state of the system. Large scale 
asymmetries in the temperature and density profiles in the hot gas in the 
cores of galaxy clusters, such as those from cold fronts and
shocks, have been used to identify supersonic mergers of subcluster 
components in those systems (e.g. see Markevitch \etal 2002, 2005). 
Such signatures would be expected in the hot IGM of groups, such as 
LGG~316, if the ongoing encounter is the supersonic collision of two 
galaxy subgroups. 

\begin{figure*}
\begin{center}
\includegraphics[height=2.47in,width=2.4in]{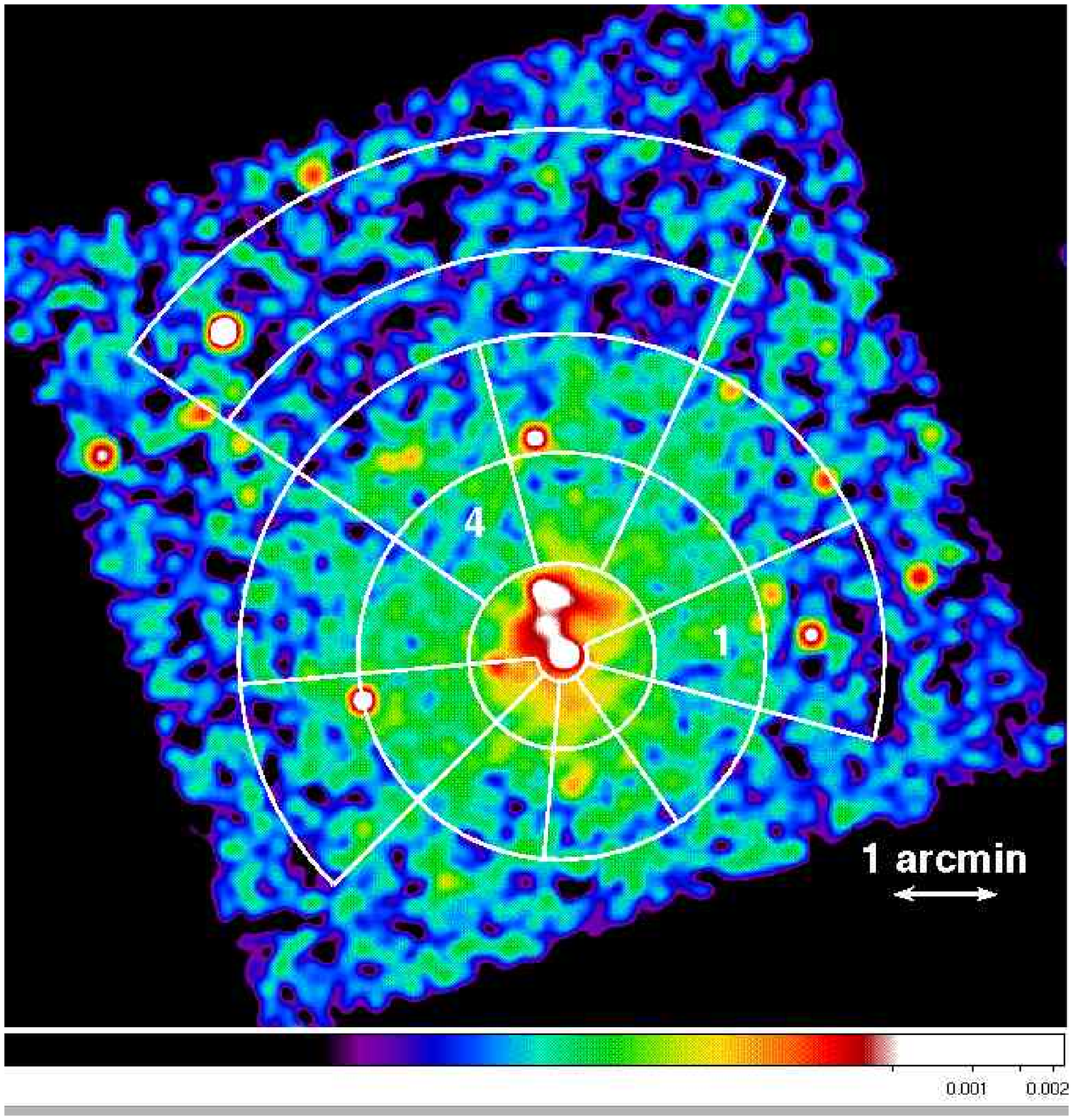}
\includegraphics[height=2.4in,width=2.4in]{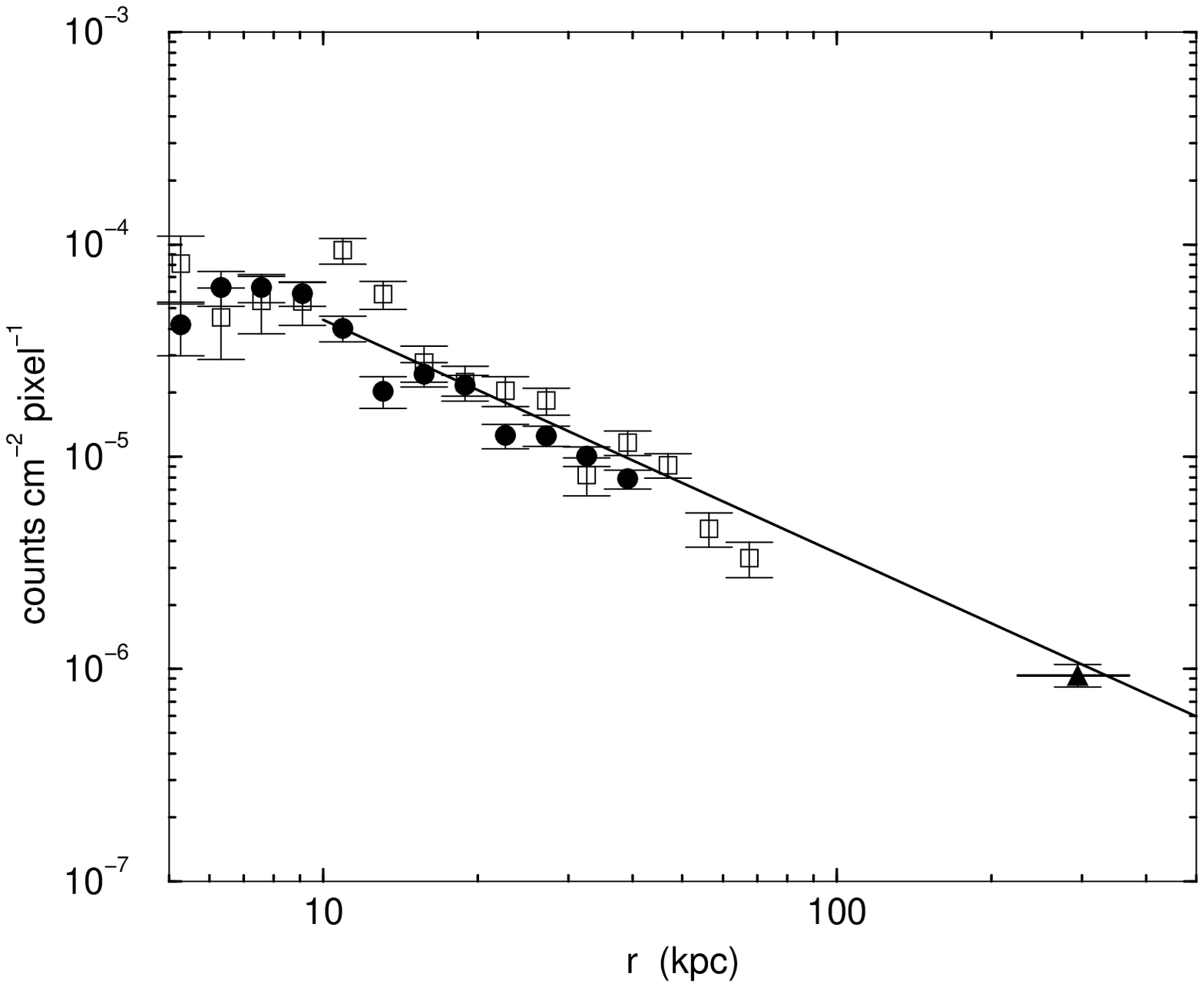}
\includegraphics[height=2.4in,width=2.4in]{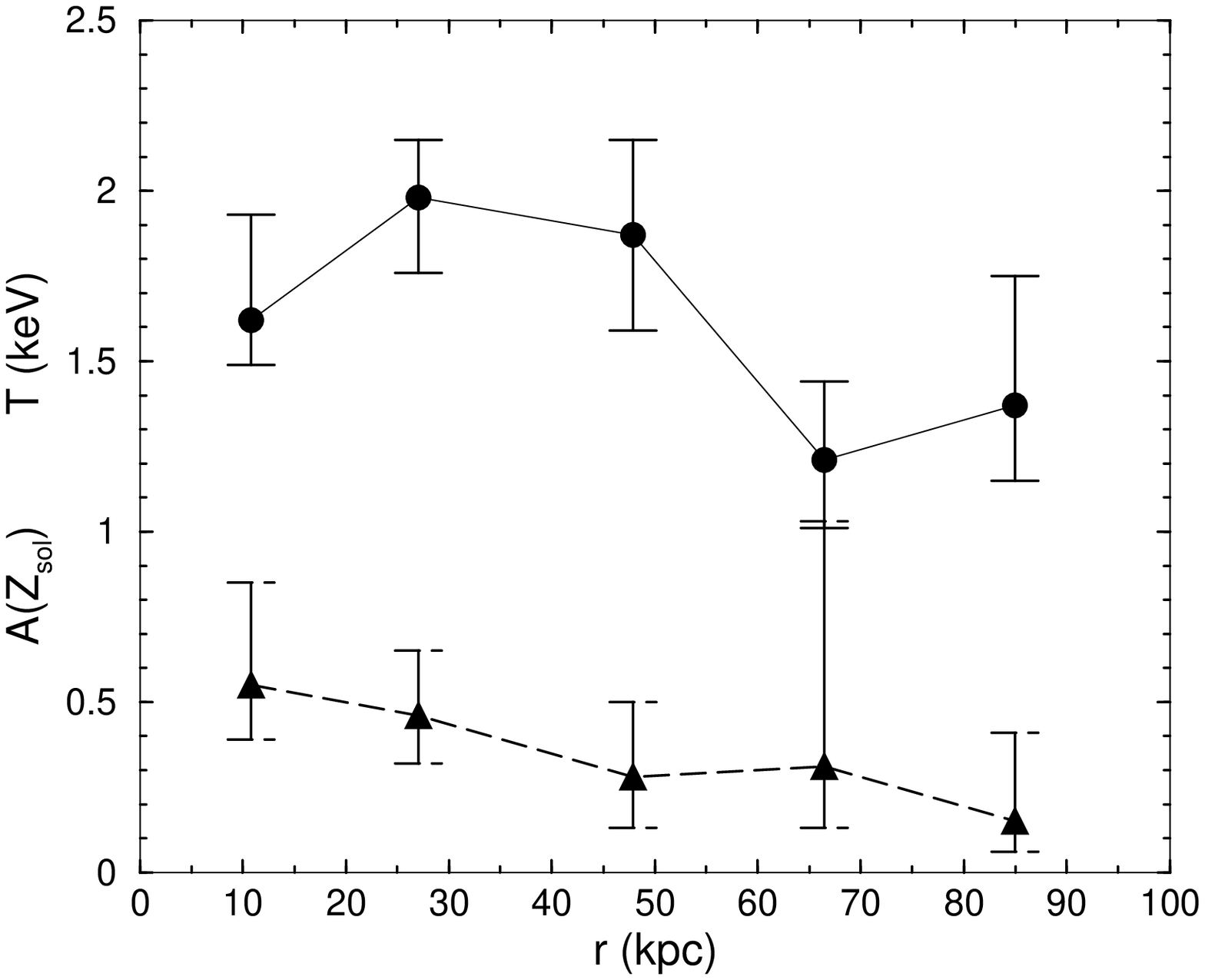}
\includegraphics[height=2.4in,width=2.4in]{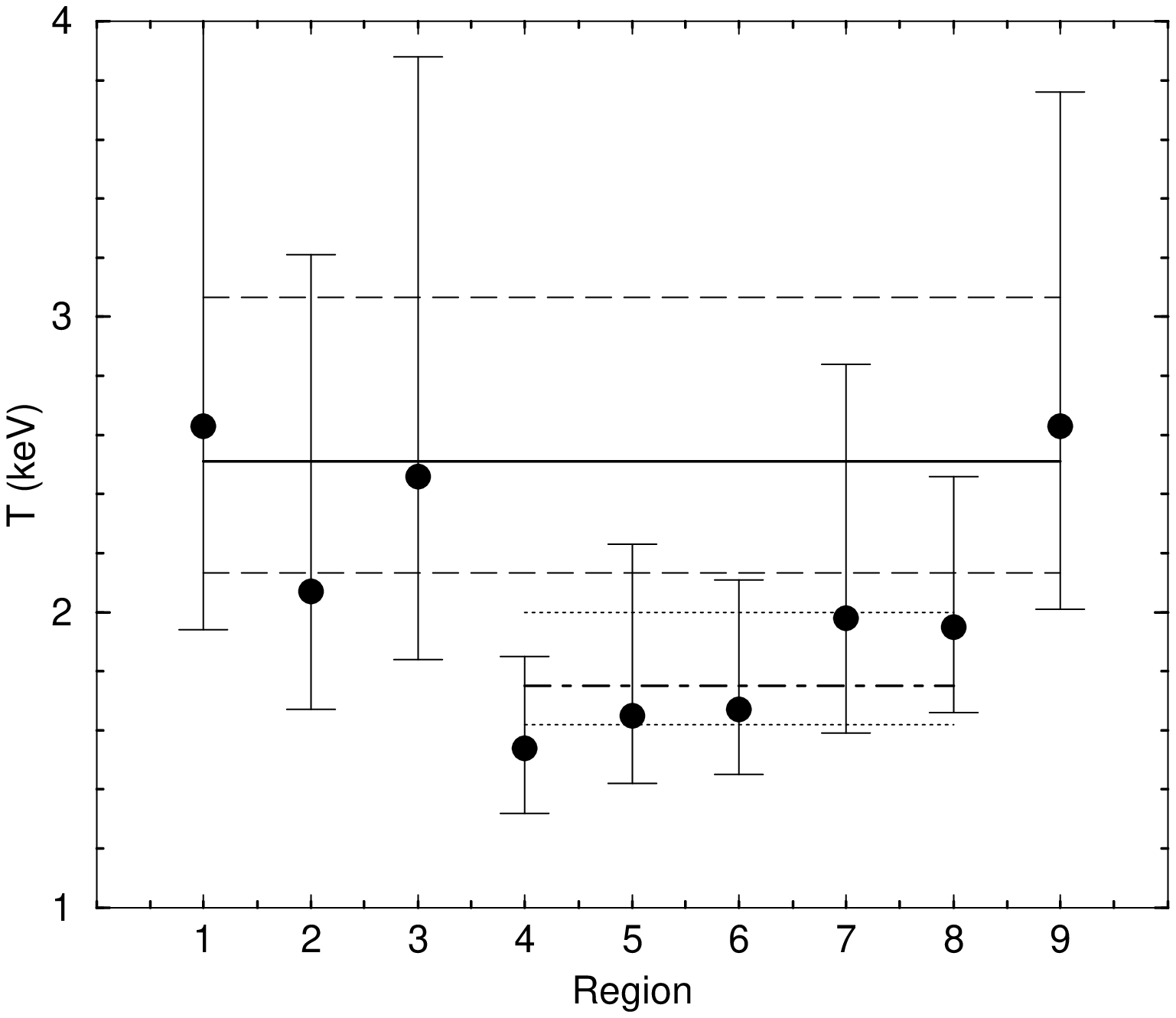}
\caption{ ({\it upper left}) $0.3-2$\,keV background subtracted, exposure
  corrected {\it Chandra} image of the region (chip S3) surrounding
  the dumbbell galaxies NGC~4782 and NGC~4783 with spectral regions 
  superposed. The image has been 
  Gaussian smoothed with a $\sigma = 4''$ kernel to highlight the 
  X-ray emission from the group IGM.  
  The spectral regions are $40^\circ$ annular sections numbered 
  counterclockwise starting from $345^\circ$. 
  Labels $4$ and $1$ mark the locations of the eastern and western radio
  lobes, respectively.
  ({\it upper right}) $1-2$\,keV surface brightness profiles as a
  function of the radial distance $r$ measured from the center of 
  NGC~4782 in an angular wedge from $38^\circ$ to $88^\circ$ 
  to the north (open squares) and from $200^\circ$ to $336^\circ$
  to the south (filled circles), measured 
  counterclockwise from west. The filled triangle denotes the mean
  surface brightness in a rectangular region on ccd S1 spanning radii 
  $ 225 \leq r \leq 370$\,kpc (horizontal line). 
  The diagonal line shows the 
  surface brightness profile expected for gas with temperature 
  (abundance) of $1.4$\,keV ($0.4\Zs$), respectively, and electron 
  density of the form $n_e = 0.0025(r/15.45\,{\rm kpc})^{-1.05}$.
({\it lower left}) IGM temperature (circles) and abundance (triangles),
 with $90\%$ confidence
level uncertainties, as a function of
  radial distance $r$ from the center of NGC~4782.
 ({\it lower right}) Gas temperature measured in the $40^\circ$ sections
for region P, shown in the upper left panel (see also 
Table \protect\ref{tab:specregsv2}), as a function of region label.
Solid and dot-dashed 
lines denote the temperature from the simultaneous fit
to the western segment (sectors $1-3$ \& $9$) and eastern segment 
(sectors $4-8$), respectively, with the $90\%$ confidence levels 
for the western (eastern) segment shown as thin dashed (dotted) lines. 
}
\label{fig:igm}
\end{center}
\end{figure*}

In the upper left panel of Figure \ref{fig:igm}, we show the
$0.3-2$\,keV image of the full ACIS S3 CCD. 
The galaxies are embedded in a roughly circular 
region of dense gas in the group's core, extending $\sim 50$\,kpc 
from the center of the dumbbell system. 
We find no visual evidence of asymmetries in the surface 
brightness outside the central interaction region, i.e. at 
$r \gtrsim 17$\,kpc, that might be expected for a 
high velocity merger of two subgroups, as suggested by Quintana 
\etal (1996).    

In the upper right panel of Figure \ref{fig:igm}, we show the surface 
brightness profile as a function of radial distance from the center
of NGC~4782 taken in two angular sectors (chosen to exclude 
regions containing the eastern and western radio lobes), with angles, 
 measured counterclockwise from west, ranging from $38^\circ$ to 
$88^\circ$ for the northern profile 
(open squares)  and from $200^\circ$ to $336^\circ$ for the southern
profile (filled circles). 
Since {\it ROSAT} All Sky Survey maps of the Galaxy show excess 
soft X-ray emission near NGC~4782/4783, we construct the radial 
surface brightness profile in the $1-2$\,keV energy band where 
this soft Galactic background is negligible. To probe the group 
emission at larger radii, we also show the surface brightness 
determined from a $443" \times 443"$ rectangular region on ccd 
S1 (filled triangle). Outside the cores of the galaxies 
and regions of enhanced emission due to stripped galaxy gas, 
i.e. radii $r \gtrsim 15$\,kpc), we find that the
surface brightness profiles both to the north and south of the
interacting galaxy pair can be described by a single power law,
consistent with gas residing in the gravitational potential of 
a single, large group. From the surface brightness we infer an 
electron density for the IGM of 
$n_{\rm IGM} \sim 2.5^{+0.3}_{-0.2} \times 10^{-3}\,
 {\rm cm}^{-3}\,(r/15.45\,{\rm kpc})^{-3\beta}$ with $\beta = 0.35$,
where the uncertainties reflect the uncertainties in the spectral 
models for the IGM gas, determined  by varying the gas temperature 
between $1.0-1.8$\,keV and metallicities between $0.3-0.5\Zs$ 
(see the lower left panel of Fig. \ref{fig:igm}). 

We first  determined the mean temperature 
and abundance of the group IGM as a function of radius by fitting 
the spectrum in five annular regions (P$_s$, P, P$_{n1}$, P$_{n2}$,
P$_{n3}$; see the upper left panel of 
Fig. \ref{fig:igm} and Table \ref{tab:pandageom}), 
covering as much of CCD S3 as possible ($r \lesssim 100$\,kpc) 
outside the interaction region contaminated by galaxy gas, and, 
unless otherwise noted, our discussion of thermal properties  
of the group IGM are based on the spectral fits to these regions.
We found excess emission in the $0.5-0.6$\,keV band in each spectrum, 
consistent with a background due to local emission from 
Galactic oxygen above that subtracted by the blank sky backgrounds. 
We modeled this soft excess background 
contribution to each spectrum by adding an additional local ($z=0$)  
low temperature APEC component, fixing all of the APEC parameters for 
the soft Galactic oxygen component at the best fit
values from a two component APEC fit to 
the spectrum of annulus P$_{n1}$. Region P$_{n1}$ was chosen
because it has good statistics ($2269 \pm 67$ net counts) and, with an 
inner radius of $37$\,kpc, is well outside the dumbbell galaxies, 
avoiding contamination by galaxy gas. We found a best fit temperature
of $0.18^{+0.03}_{-0.04}$\,keV and metallicity of $1.2 (> 0.2)\,\Zs$ 
for the soft Galactic oxygen background at $z=0$, and a 
temperature and metal abundance of $1.87 \pm 0.28$\,keV and 
$0.28^{+0.22}_{-0.15}\,\Zs$ ($\chi^2/{\rm dof} = 121/133$) ,
respectively, for the LGG~316 group IGM ($z=0.0154$) in that region.
We found a mean $0.3 - 2$\,keV surface brightness from the 
soft Galactic component of  $5.7 \times 10^{-19}$\ergscm\,arcsec$^{-2}$.

We checked that this model for the local soft X-ray component was
robust in two ways. First, since the determination of the abundance 
for the local emission is uncertain, we varied the abundance value for 
the local emission in region  P$_{n1}$ from $0.2-2\Zs$ and found that the 
temperature of the soft Galactic component was unchanged 
(at $90\%$ confidence limits) and the $0.3-2$\,keV fluxes for the soft 
Galactic component varied by $\lesssim 10\%$. 
Second, assuming that the mean flux per pixel in the soft Galactic
component is constant across the detector, we used our model based on
region P$_{n1}$ to predict the $0.3-2$\,keV soft Galactic flux expected 
from a $443" \times 443"$ rectangular region on ccd
S1, located $\sim 15'.6$ from NGC~4782, where the 
ratio of soft Galactic emission to group IGM emission is $\sim 2.3$. 
We then fit the spectrum of this rectangular region with a two 
component APEC model, fixing the temperature and abundance 
($0.18$\,keV and $1.2\Zs$, respectively) from our  P$_{n1}$ model for the 
soft X-ray background, but allowing the model normalization to vary. 
Metal abundances of the second APEC (IGM) component were varied
between $0.1-0.3\Zs$, with the temperature and normalization
determined from the fit. In all cases, the measured  
$0.3-2$\,keV soft Galactic foreground flux from this region on ccd S1 
differed by less than $6\%$ from that predicted by our  background model 
rescaled from region  P$_{n1}$. Similarly, the temperatures 
found for the second 
APEC component ($\sim 1-2$\,keV) were consistent with emission from
the group IGM. 

Thus we fix the normalization of the soft Galactic background
component for all other spectral regions at the best fit value for region 
P$_{n1}$ scaled by the ratio of the region areas. 
We then fit the spectra for the annular regions, allowing the
temperature, abundance and normalization of the IGM APEC component to
vary. Our results are listed in Table \ref{tab:igmfitswO}.
In the lower left  panel of Figure \ref{fig:igm}, 
we plot the mean temperature
(circles, solid line) and mean abundance (triangles, dashed line) 
from these model fits as a function of radial distance from NGC~4782. 
Within the uncertainties, the group gas is nearly isothermal at
$\sim 1.4 \pm 0.4$\,keV. 
The measured metal abundances of 
$\sim 0.3-0.5\,\Zs$ are constant at the $90\%$ confidence level. 

To investigate  any angular change in the temperature  
of the surrounding group IGM, we divide each annulus into 
sectors (see the upper left panel of Fig. \ref{fig:igm}), 
based on nine $40^\circ$ 
sectors of the complete reference annulus P.
 The annular sectors in region P were chosen
such that the western radio lobe 
lies mainly in annular sector $1$ and the eastern radio lobe in
annular sector $4$. For annulus P, 
we fit an absorbed single temperature 
APEC model to the 
IGM emission in the individual annular sectors, 
fixing the abundance at the best fit value ($0.46\Zs$) for the 
annulus as a whole. Our results are plotted in the lower right panel 
of Figure \ref{fig:igm}.
The  data suggest that the IGM gas 
temperature may be higher to the west than to the east in this 
annulus ($17 \leqslant r \leqslant 37$\,kpc) . To test this we 
divided the annulus P into two segments, the 
eastern segment (annular sectors $4-8$) and the western segment 
(annular sectors $1-3$ and $9$), and simultaneously fit the spectra
of the annular sectors in each segment, using an absorbed APEC model 
as above. Our results, shown as the dot-dashed and solid 
lines for the eastern 
and western segments, respectively, in the lower right panel of
 Figure \ref{fig:igm}
show that, at the $90\%$ confidence level, the mean temperature of 
the gas in the western segment 
is higher 
($kT = 2.51^{+0.55}_{-0.38}$\,keV) than the IGM gas 
temperature in the eastern segment at the same radii 
($kT = 1.75^{+0.25}_{-0.13}$\,keV). From fitting the spectra of the 
angular sectors in the remaining outer annuli , we find no evidence 
(at $90\%$ confidence) for   
angular variation in the IGM gas temperature at these larger radii 
($r > 37$\,kpc).
Thus the overall gas morphology and temperature of the IGM for radii 
$r\gtrsim  37$\,kpc is 
consistent with a single group of galaxies, in which the group gas  at large 
radii is expected to be close to isothermal and in hydrostatic 
equilibrium; while the increased temperature to the west, close to the 
interaction region, may be evidence for the recent supersonic passage of
NGC~4783 to the west of NGC~4782, in agreement with simulations 
(Borne \etal 1988; Madejsky \& Bien 1993), thereby heating the IGM.

\subsubsection{Mass, Luminosity and Group Scaling Laws}
\label{sec:groupscale}
 
We use the mean temperature of the group gas in the 
outer annuli in the M-T relation from 
Finoguenov \etal (2001) to estimate the mass of the group, 
\begin{equation}
M_{500} = (2.64^{+0.39}_{-0.34} \times 10^{13})(kT)^{1.78 \pm 0.1}\,\,.
\label{eq:mt}
\end{equation} 
From Equation \ref{eq:mt} we find a group mass of 
$4.8^{+2.7}_{-2.2} \times 10^{13}\Ms$ for 
a mean temperature $kT = 1.4 \pm 0.4$\,keV,
in agreement with the group mass determined from the ($447$\kms) 
velocity dispersion for the LGG~316 group taken as a whole.
The total X-ray luminosity of the group is more difficult to
determine, since we can only directly measure the surface brightness
profile and verify our model for the gas density to radii of 
$\sim 50$\,kpc to the south and $\sim 100$\,kpc to the north of
NGC~4782, i.e. $0.08r_{500}$ and $0.15r_{500}$, respectively. 
We determine the bolometric X-ray luminosity within radius $r$ 
by using the $1-2$\,keV flux, found from integrating the 
$\beta$-model for the surface brightness profile in Fig. \ref{fig:igm},
to compute the unabsorbed $0.01-14$\,keV flux, using APEC spectral 
models with gas temperatures $1.4 \pm 0.4$\,keV and abundances 
$0.3 \pm 0.2 \Zs$ from our spectral fits to the IGM gas. 
We find the X-ray luminosity of the group gas within a radius of 
$50$($100$)\,kpc of NGC~4782 to be 
${\rm log}\,L_{\rm X} = 41.71^{+0.14}_{-0.04}$
($41.99 ^{+0.13}_{-0.04}$)\ergs, where the uncertainties in the 
luminosity reflect the uncertainties in the spectral model
temperatures and abundances. 
However, as the upper right panel of Figure \ref{fig:igm} 
shows, data suggest that the $\beta = 0.35$ model for the X-ray surface
brightness (and gas density) remains valid to $r \gtrsim 300$\,kpc 
($\gtrsim 0.45r_{500}$), implying a lower bound on the X-ray
luminosity for the group of 
${\rm log}\,L_{\rm X} > 42.42^{+0.13}_{-0.05}$\ergs. 
Although the extrapolation of the surface brightness beyond 
$r \sim 300$\,kpc may introduce large uncertainties, we use this
extrapolation to estimate the total X-ray luminosity expected within
$r_{500} = 660$\,kpc for comparison with X-ray scaling 
relations. We find ${\rm log}\,L_{\rm X} = 42.73^{+0.13}_{-0.05}$\ergs 
for the LGG316 group. This result is in excellent agreement with 
that expected from the  $L_{\rm X}-T_{\rm X}$ and 
$L_{\rm X}-\sigma_{\rm v}$ scaling relations for a single large 
X-ray group of temperature $1.4 \pm 0.4$\,keV and 
velocity dispersion $\sim 447$\kms 
(see Osmond \& Ponman 2004, Figs. 13 and 15, respectively).

\section{X-ray Features Associated with the Radio Jets and Lobes}
\label{sec:cavities}

\begin{figure}[t]
\includegraphics[height=2.2in,width=3in]{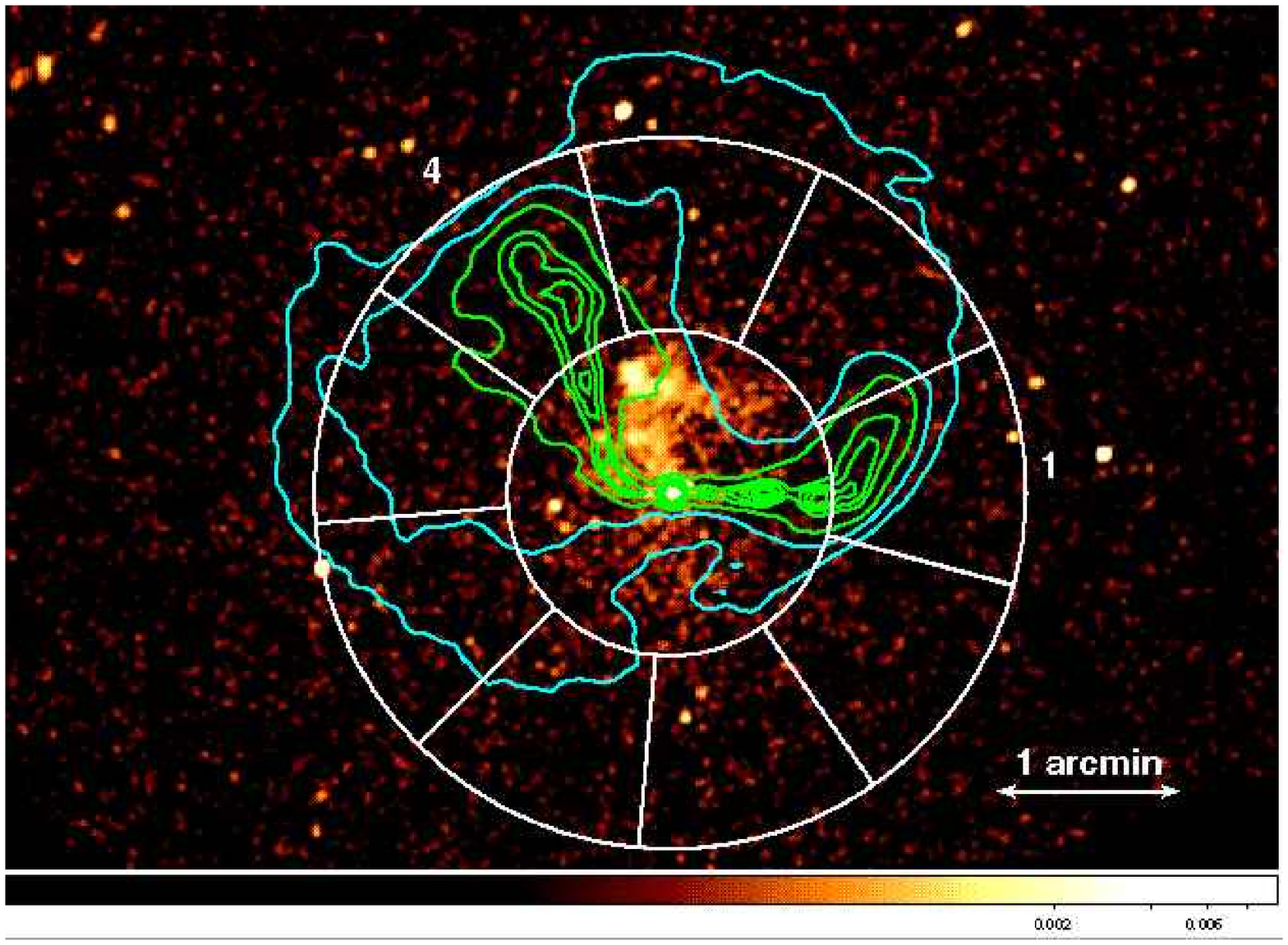}
\includegraphics[height=2.3in,width=2.3in]{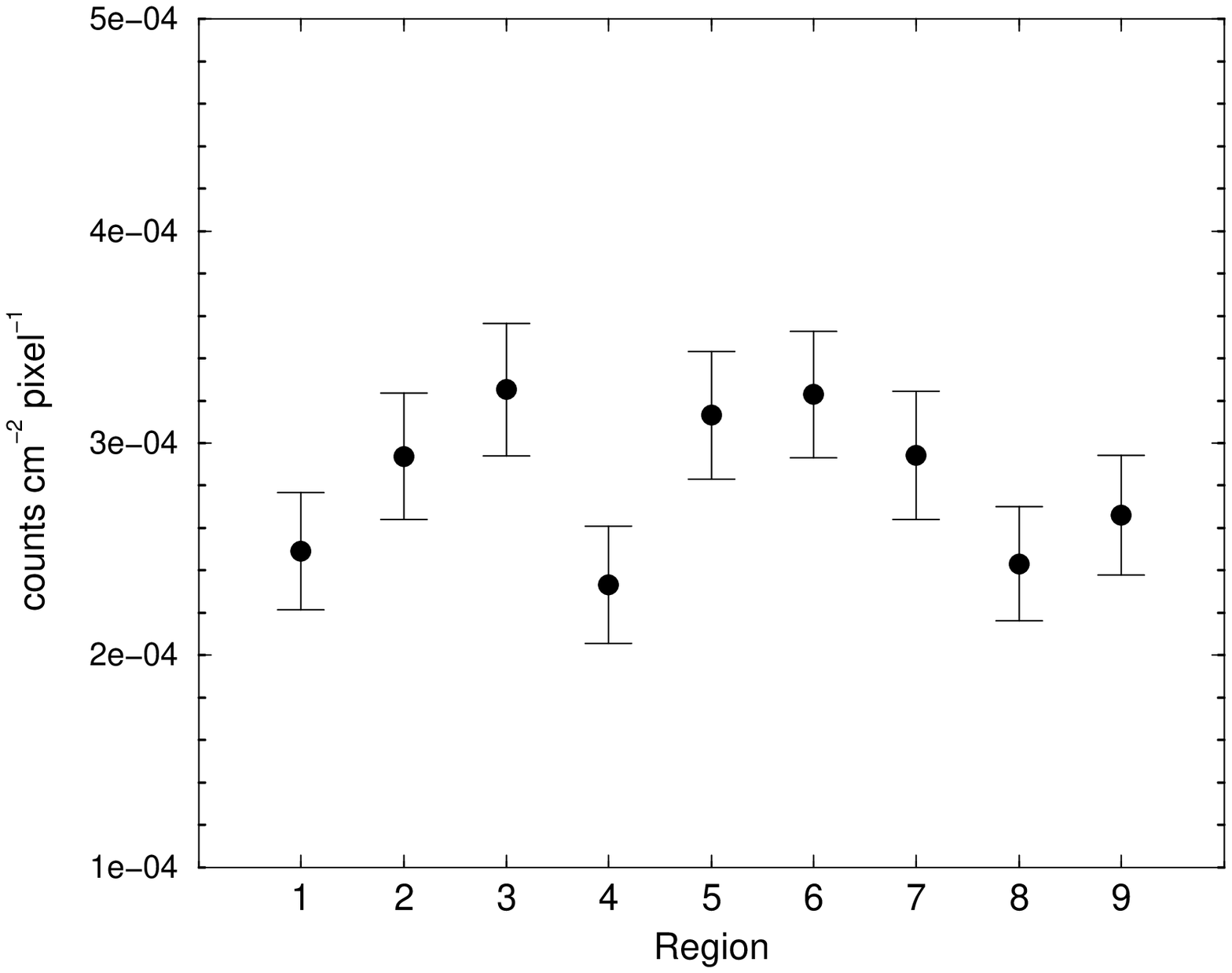}
\caption{({\it top}) $0.3-2$\,keV {\it Chandra} image with annular sectors
  from region P (labeled counterclockwise from annular sector $1$) and 
 $1.5$\,GHz VLA radio contours for NGC~4782(3C278) overlaid. The
  radio map has $6.0 \times 6.0$\,arcsec resolution with contour levels
  of $1.5$, $7$ (cyan), $15$, $25$, $30$, $35$, $40$ (green)\,mJy/beam, 
 showing the 
  radio jets and lobes (green) embedded in a faint, large-scale 
  radio halo (cyan).
  ({\it bottom}) Mean surface brightness
  in the $0.3-1$\,keV band as a function of 
  sector number for region P with mean radius 
  $ r = 27$\,kpc. Errors are $1\sigma$. The deficit in X-ray emission in 
  annular sector $4$, coincident with the eastern radio lobe, indicates
  the presence of an X-ray cavity. For comparison, the mean surface 
  brightness measurements for annular sectors $2-6$ at mean radius 
  $r=48$\,kpc (annulus $P_{n1}$) outside the
  eastern lobe, i.e.  $2.2 \pm 0.2$, $2.6 \pm 0.2$, $2.2 \pm 0.2$, 
  $2.0 \pm 0.2$, $2.0 \pm 0.2$, respectively, in units of
  $10^{-4}$\sbunit, agree within their
  $1\sigma$ uncertainties and thus show no decrement in X-ray emission 
  between region $4$ outside the radio lobe and neighboring regions. 
  $1$ pixel = $1.968'' \times 1.968''$.
  }
\label{fig:cavity}
\end{figure}

In the top panel of Figure \ref{fig:cavity}, we superpose contours from 
a $1.5$\,GHz VLA map on the $0.3-2$\,keV Chandra image to show the 
correspondence between the large scale radio structure of 3C\,278 and 
X-ray emission from NGC~4782/4783 and the LGG316 group. The appearance  
of the eastern and western radio jets and lobes  are very different. 
The eastern jet shows a prominent bend towards the northeast at 
$\sim 5-6$\,kpc ($17-18''$) from the nucleus of NGC~4782. 
Shortly after the bend, the eastern jet inflates into an approximately 
spherical lobe of radius $\sim 10.7$\,kpc ($ 34''$). The western jet 
extends straight for $\sim 12.6$\,kpc ($ 40''$) before flaring 
into the western radio lobe. The western lobe appears distorted into an 
ellipsoidal shape, bending gradually to the northwest, with the radio 
contours on its southern side compressed relative to those on its 
northern side. 
The radio jets and lobes are embedded in a faint, large-scale radio halo 
extending towards the northeast. 

\subsection{The Eastern X-ray Cavity}
\label{sec:eastcav}

We use  the individual annular sectors from the upper left panel of
Figure \ref{fig:igm}  
to search for X-ray cavities  associated with the radio lobes. 
In the annulus (P) extending $17 < r < 37$\,kpc from the center of 
NGC~4782 and shown 
in the top  panel of Fig. \ref{fig:cavity},  
annular sector $1$ contains the western 
radio lobe, while the eastern radio lobe almost entirely fills the  
annular sector $4$. In the bottom panel of Figure \ref{fig:cavity}, 
we plot the 
mean surface brightness in the $0.3-1$\,keV energy band for 
each annular sector of region P as a function of sector number. 
For comparison, we also list the mean surface brightness for annular 
sectors $2-6$ at larger radii ($37 < r < 59$\,kpc) from region 
$P_{_n1}$ outside  the eastern radio lobe, i.e. annular sector $4$ 
and the two neighboring annular sectors on either side.  
We find a $29^{+12}_{-13}\,\%$ decrease ($1\sigma$ uncertainties) in the 
mean surface brightness in sector $4$ of annulus P,
at the location of the eastern radio lobe, while there is no analogous 
decrease in mean surface brightness 
in annular sector $4$ of region P$_{n1}$ outside the eastern lobe,  
compared to its neighboring annular sectors. This  
indicates an X-ray cavity coincident with the eastern radio lobe.
We do not observe a statistically significant decrease in 
the surface brightness ($\sim 11^{+19}_{-11}\,\%$, $1\sigma$
uncertainties) in the annular sector $1$ of annulus P,
that contains the western radio lobe. 

The expected reduction in X-ray  
surface brightness is small for cavities whose inclination angle 
is large, so that X-ray cavities are difficult to observe unless the cavity 
lies close to the plane of the sky. 
We define the surface brightness decrement as the 
ratio of the observed surface brightness in the cavity to that
expected if there is no cavity, and find the observed decrement in 
the eastern lobe is $0.71^{+0.13}_{-0.12}$ ($1\sigma$ uncertainties). 
Assuming that the radio plasma in the eastern lobe has emptied the 
cavity of X-ray emitting gas, we can use the reduction in X-ray surface 
brightness to constrain the inclination angle of the eastern radio lobe with
respect to the sky. Modeling the expected surface
brightness decrement in the $0.3-1$\,keV 
band by integrating through lines of sight passing through a
spherical cavity coincident with the eastern radio lobe, as a function 
of the inclination angle of the lobe, 
we find that the observed decrement in 
the eastern lobe suggests that the eastern lobe is inclined at an angle of 
$\lesssim 30^\circ$ with respect to the plane of the sky. 

\subsection{Jet Emission in X-rays}
\label{sec:jetemiss}

\begin{figure}[t]
\begin{center}
\includegraphics[height=2.23in,width=3in]{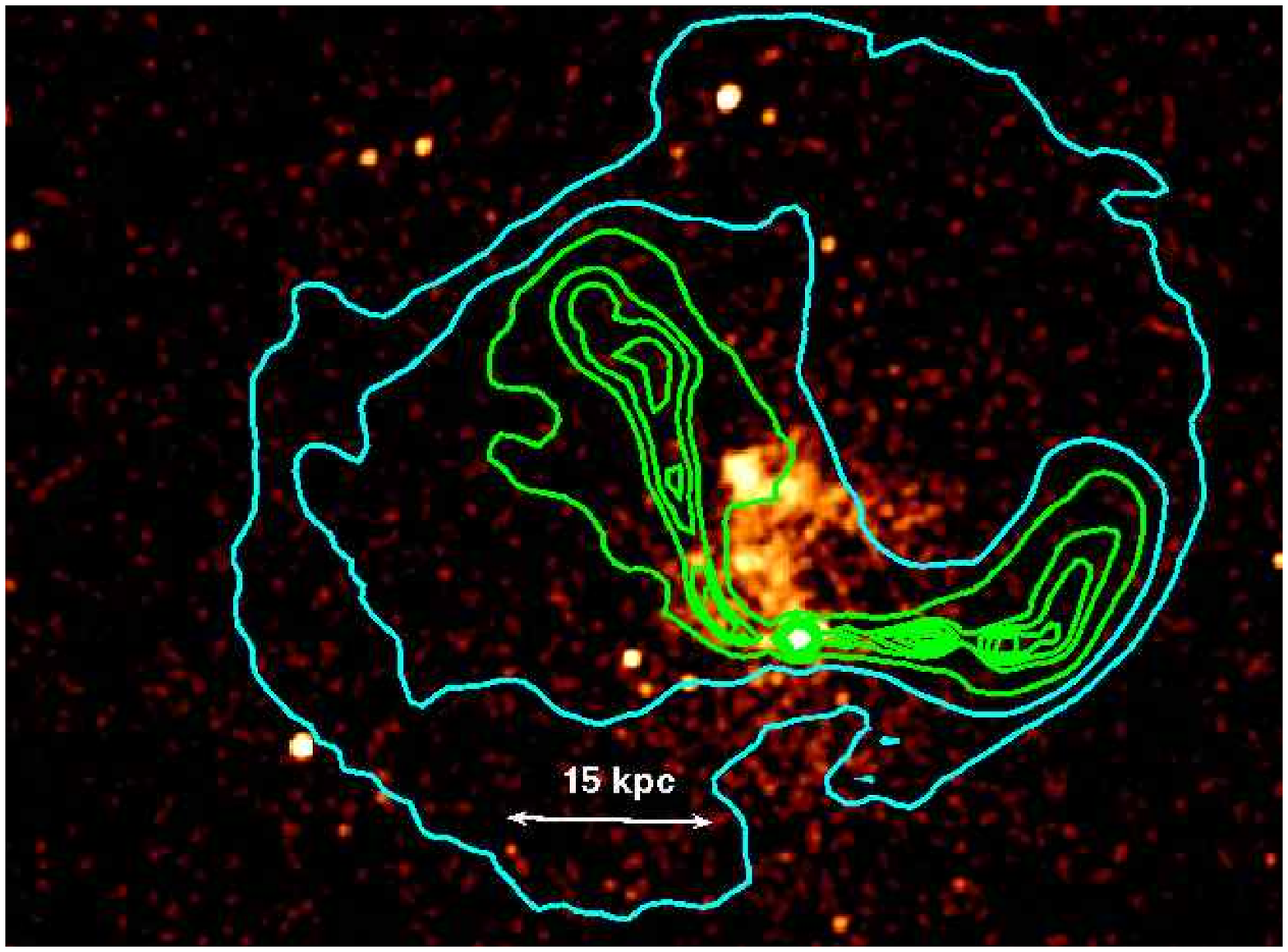}\vskip 0.0in
\includegraphics[height=2.23in,width=3in]{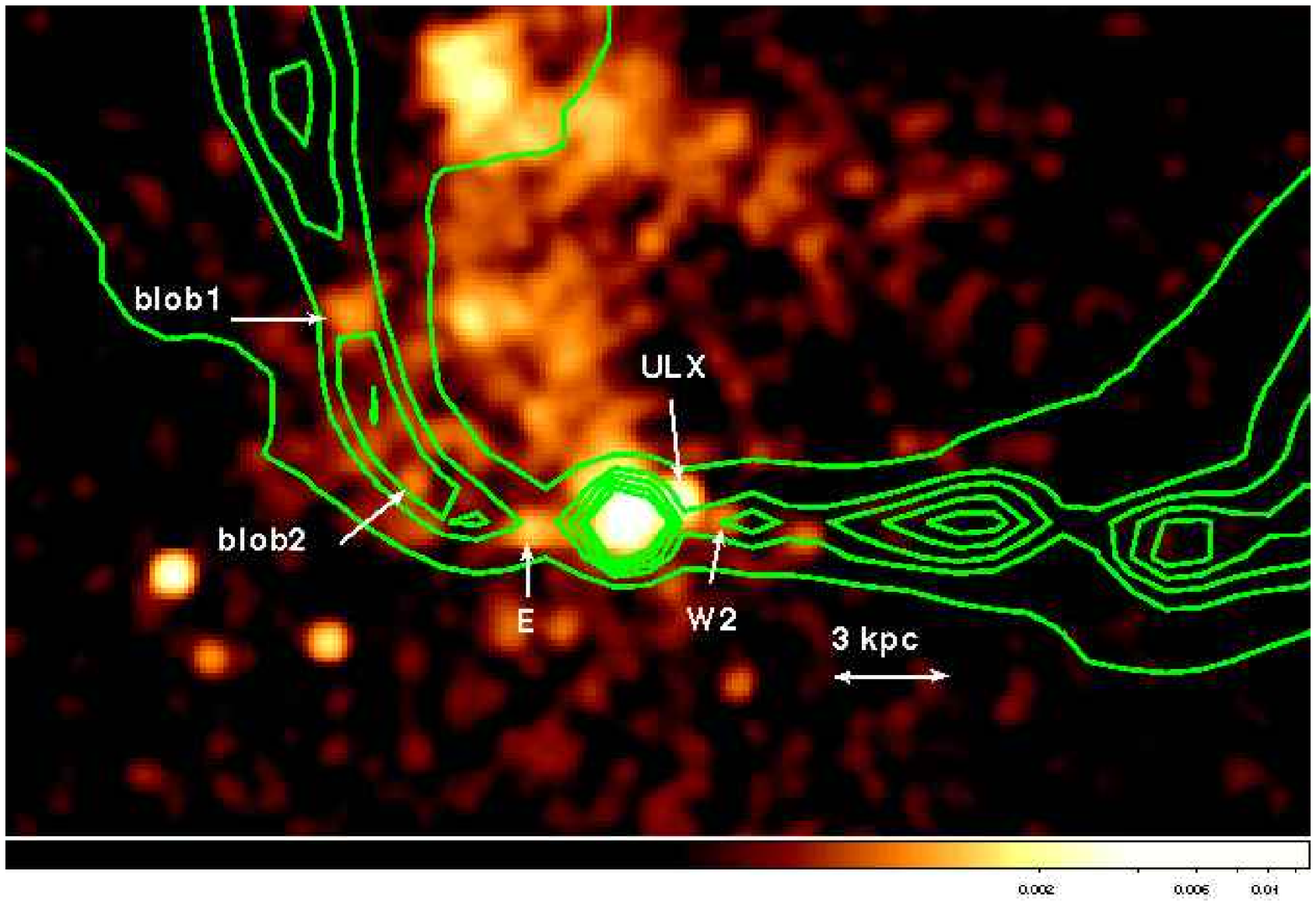}\vskip 0.0in
\includegraphics[height=2.23in,width=3in]{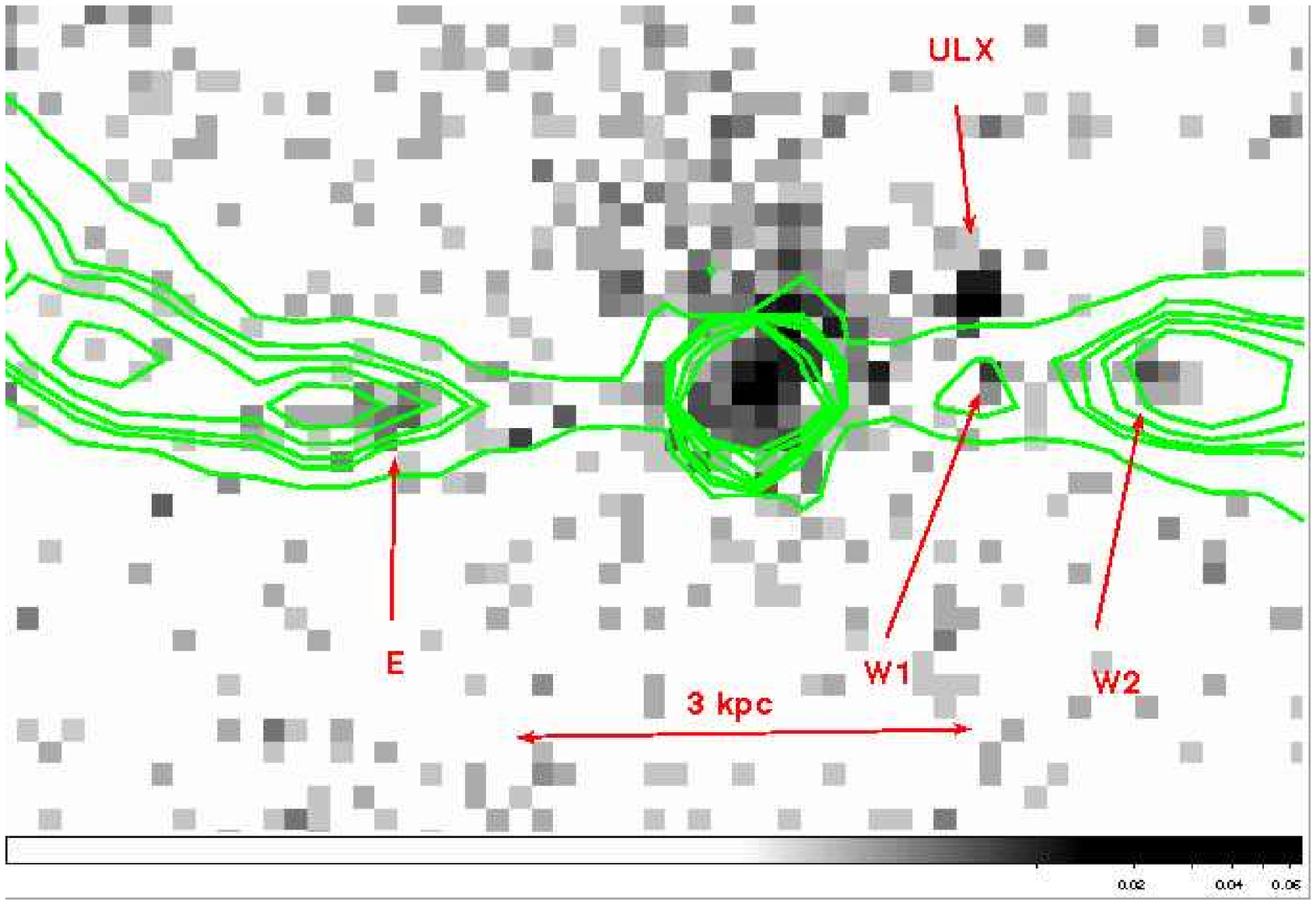}
\caption{({\it top}) $0.3-2$\,keV {\it Chandra} image with 
$1.5$\,GHz VLA radio contours for NGC~4782(3C278) as in
  Fig. \protect\ref{fig:cavity} overlaid showing  the large scale radio
  structure of 3C178 for reference. The X-ray image has been
  background subtracted, exposure corrected and smoothed with a $1''$
  Gaussian kernel. ({\it middle}) Zoom-in of the 
same image as above. `Blob1(2)' identifies extended emission in the 
X-ray `fan' possibly associated with the eastern radio jet.  
({\it bottom}) $0.3-2$\,keV {\it Chandra} image of the inner $\sim
5$\,kpc of NGC~4782 with contours from a $1.0 \times 1.0$ arcsec resolution 
$4.9$\,GHz VLA radio map of the inner jets of 3C278 overlaid. 
Radio contour levels are $0.2$, $0.4$, $0.5$, $0.7$, $1.0$, $2.0$, and 
$3.0$ mJy/beam. E and W1, W2 label X-ray point sources (knots)
coincident with the end of the inner eastern jet and the double peaked 
structure of the western radio jet, respectively.  The ULX north of 
NGC~4782's nuclear region is also labeled.
}
\end{center}
\label{fig:radioknots}
\end{figure}
In Figure 5
we show the correspondence between the 
X-ray emission and the radio structure of 3C278 in more detail. 
We find X-ray emission associated with each of the two radio jets. The 
top panel shows (as in the top panel of Fig. \ref{fig:cavity}) the 
large scale radio structure from $1.5$\,GHz VLA measurements 
superposed on the $0.3-2$\,keV Chandra image for reference. In the 
middle panel, we see that the eastern radio jet bends sharply 
through the `fan' of X-ray
emission extending northeast of NGC~4782, with $\sim 60\%$ of the 
$0.3-2$\,keV `fan' emission from the region lying, in projection, 
within the eastern jet. The brightest 
emission region in the fan (labeled `blob1' in the middle panel of 
Fig. 5 )
is located near the base of the eastern radio 
plume, and  contains $\sim 14 \%$ ($34 \pm 6$ source counts 
within a 3'' circular region) of the total $0.3-2$\,keV fan emission.  
We also find $11 \pm 3$ source counts in the $0.3-2$\,keV band in a 
$1.5''$ circular region (blob2) near the bend in the eastern jet.
If these `blob' regions are associated with the radio jet, 
we may be observing directly its  deceleration and/or jet-plume
transition.
 
In the lower panel of Figure 5 
we show a $0.3 - 2$\,keV {\it Chandra} image of the inner $\sim 5$\,kpc 
region surrounding the nucleus of NGC~4782 with radio contours from a 
 $1'' \times 1''$ resolution $4.9$\,GHz VLA map of the same 
region superposed. Two sources, labeled E and W2, are coincident 
with the radio bright ends of both the inner eastern and western jets. 
We estimate the background
rates for these two sources using a local annulus surrounding each
source. We find that these sources are soft,   
with net source counts in the $0.3-2$\,keV band of $16.5 \pm 4.0$ and 
$10.7 \pm 3.3$ ($1 \sigma$ uncertainties) and in the $2-8$\,keV band
of $0.8 \pm 1.5$ and $0.1 \pm 1.0$ ($1 \sigma$ uncertainties), 
respectively. Assuming Galactic absorption, these colors   
are consistent with steep photon indices, i.e. $\Gamma \gtrsim 2.0$. 
Using the observed count rates and 
this power law spectral model, we estimate $1$\,keV flux densities 
of $\lesssim 0.5$\,nJy and $\lesssim 0.4$\,nJy for the E and W2 
sources. From the VLA radio map, we measure  peak $4.9$\,GHz radio 
fluxes of $4$ and $6$\,mJy for the 
E and W2 sources, respectively, and estimate 
$2$-point spectral indices of $\alpha \gtrsim 0.9$. Although the 
uncertainties are large due to our low statistics, these values are 
consistent with that expected for synchrotron emission from X-ray
knots due to the deceleration of the inner radio jets (see, e.g. 
Hardcastle \etal 2005b). We also find X-ray emission ($12 \pm 3.5$ counts 
when $3 \pm 0.4$ background counts are expected) extending along the 
axis of the inner western jet, with 
a knot of X-ray emission  (labeled W1 in Fig. 5) 
coincident with a second peak in the $4.9$\,GHz radio emission in 
this region. 

\section{Constraining the Interaction Kinematics}
\label{sec:velcon}

With the thermodynamic properties of gas in the NGC~4782/4783 
dumbbell and in the surrounding group 
determined, we use these properties to constrain the 
kinematics of the interactions. 
As shown in Figure \ref{fig:chandra}, the northern galaxy NGC~4783 shows
a sharp surface brightness discontinuity (``cold front''; see Vikhlinin
\etal 2001; Markevitch \etal 2000) on its eastern
side. In \S\ref{sec:edges} we 
use the gas properties on either side of this surface brightness
edge to constrain the three-dimensional velocity of 
NGC~4783. 
The evolution of radio jets and lobes can be strongly influenced by
ram pressure from the surrounding gas, caused by either bulk motions 
in the ISM/IGM  or by the motion of their host galaxy through 
the intragroup medium (see, e.g. Eilek \etal 1984).  
In \S\ref{sec:bend} and \S\ref{sec:cavconst} we determine  
constraints on the ram pressure velocity of gas impacting the 
radio jets and lobes, inferred from the thermodynamic properties of
the surrounding gas, the jet and lobe 
morphology and the properties of the X-ray cavity  
associated with the eastern radio lobe.  

\subsection{Cold Front Analysis Constraints on NGC~4783's Velocity}
\label{sec:edges}

\begin{figure}[t]
\includegraphics[height=3in,width=3in,angle=270]{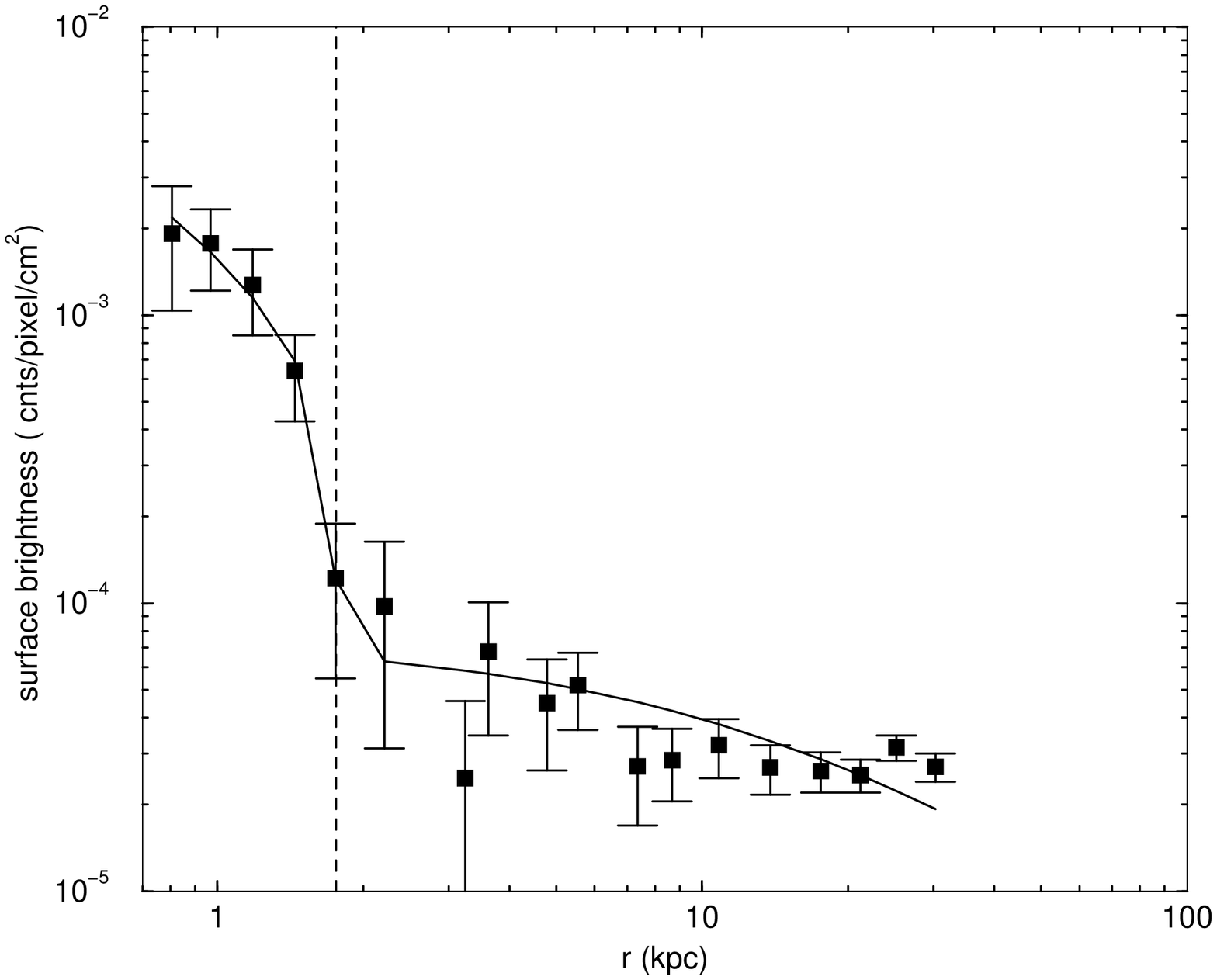}
\includegraphics[height=3in,width=3in,angle=270]{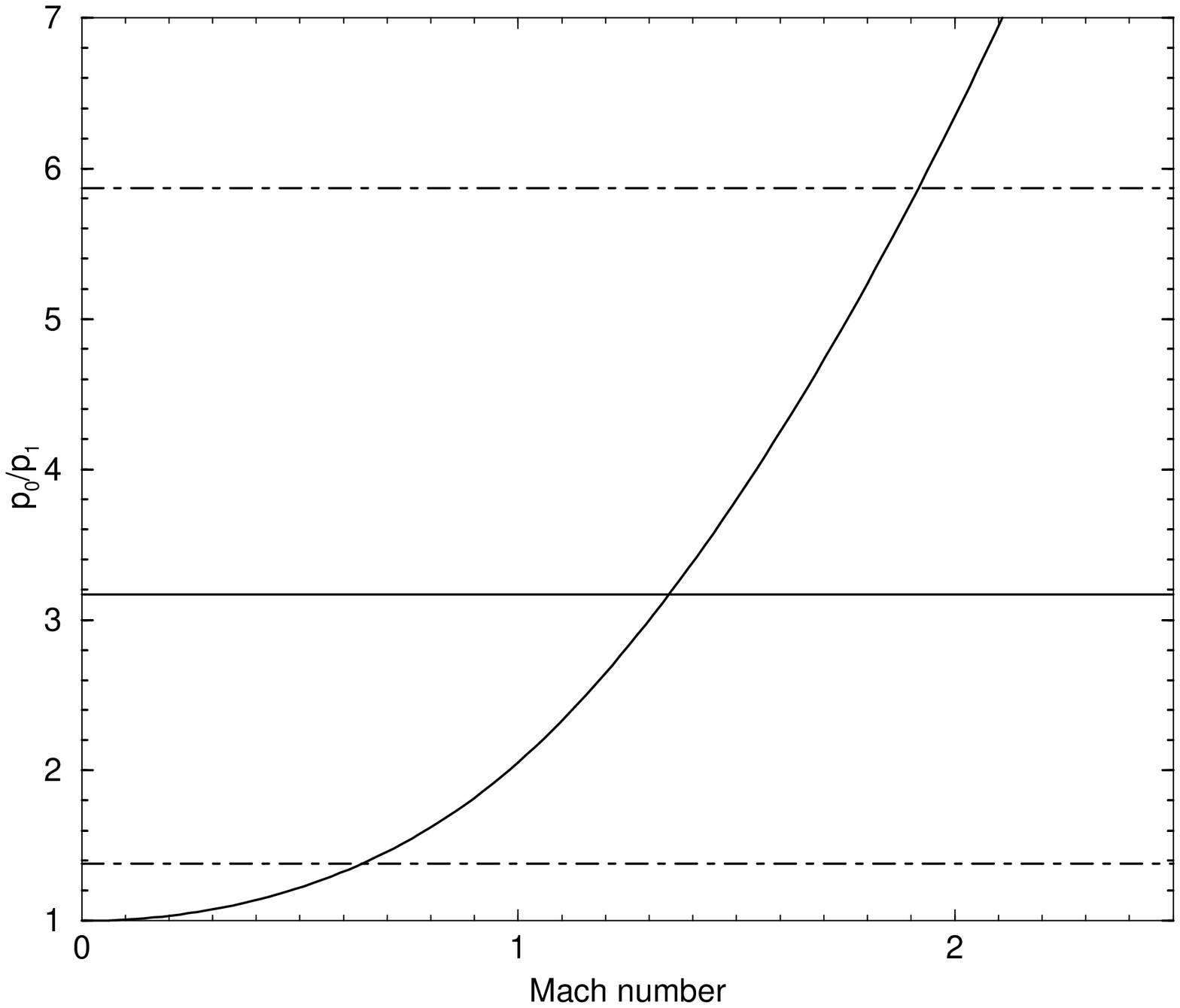}
\caption{({\it top}) $0.3-2$\,keV surface brightness profile
  from elliptical annuli concentric to the edge
  in NGC~4783 and constrained to lie in the angular sector between 
$94^\circ$  and $186^\circ$ (measured counterclockwise from west) as a 
function of the distance from the nucleus of NGC~4783. 
The vertical dashed line denotes the 
best fit position of the edge ($r_{\rm edge} = 1.76$\,kpc). 
The solid line denotes the density model fit to the surface 
brightness for the cavity-uncorrected data (Model A)
assuming $n_{\rm gal} \propto (r/r_{\rm edge})^{-\alpha_1}$ for galaxy 
gas inside the edge ($r < r_{\rm edge}$) and  
$n_{\rm IGM} = K(r_{\rm IGM}/15.45\,{\rm kpc})^{-3\beta}$, with
$\beta=0.35$ and $r_{\rm IGM}$ the distance from the center of
  NGC~4782,  
for $r > r_{\rm edge}$. 
({\it bottom}) Pressure ratio between the undisturbed IGM and pressure 
inside the edge versus Mach number of the galaxy from 
Vikhlinin \etal (2001).  The solid 
horizontal line denotes the pressure ratio derived from 
the model fit shown in the left panel. The 
dot-dashed horizontal lines denote the uncertainty in the pressure ratio, 
due to the $90\%$ CL uncertainties in the galaxy and IGM temperatures and 
the IGM abundance, and due to variation of the galaxy metallicity 
over the $0.3-1.2\Zs$ range.
}
\label{fig:n4783edgefit}
\end{figure}
Following the method of Vikhlinin \etal (2001), we define a bounding 
ellipse centered on the NGC~4783
nucleus with semi-minor (semi-major) axes of 
$4''.5$ ($8''.8$) and position angle $304^\circ$,  
that traces the surface brightness edge with the 
minor axis oriented along NGC~4783's direction of motion in 
the plane of the sky. We then construct the surface brightness 
profile, shown in Figure \ref{fig:n4783edgefit}, from 
elliptical annuli concentric to the bounding ellipse and 
constrained to lie in an angular sector between  $94^\circ$  
and $186^\circ$ (measured counterclockwise from west). 
This sector was chosen to exclude the `horn' of emission extending 
northwest from NGC~4783 and emission from the bridge 
south of NGC~4783. In Figure \ref{fig:n4783edgefit}, 
the radius $r$ is the weighted mean for each annular 
region measured from the center of NGC~4783 outward along NGC~4783's 
projected direction of motion. 

We fit the surface brightness profile across the edge by integrating simple 
spherically symmetric density models along the line of sight as a 
function of the projected distance $r$.
We assume a power law model for the density distribution,   
$n_{\rm gal} = n_{\rm in}(r/r_{\rm edge})^{\alpha_1}$,  
for galaxy gas inside the edge ($r \le r_{\rm edge}$). 
Outside the edge, we assume 
that the group gas density is well represented by  
the power law fit to the IGM surface brightness profile from  
\S\ref{sec:grouphydro}, i.e. 
$n_{\rm IGM} = K(r_{\rm IGM}/15.45\,{\rm kpc})^{-3\beta}$ with
$\beta=0.35$ from our previous analysis and 
$r_{\rm IGM}$ measured from the center of NGC~4782. 
The discontinuity ($J^2$) in the X-ray emission at the position of the 
surface brightness edge ($r_{\rm edge}$), is then 
\begin{equation} 
J^2 = \frac{\Lambda_{\rm in} n_{\rm in}^2}{\Lambda_{\rm out}n_{\rm  out}^2} 
\label{eq:sbjump}
\end{equation}   
where $n_{\rm in}$ ($n_{\rm  out}$)  are the gas densities and 
$\Lambda_{\rm in}$ ($\Lambda_{\rm out}$) are the X-ray emissivities  
for the galaxy (IGM) gas inside (outside) the surface
brightness edge, respectively.  
We then allow the edge position ($r_{\rm edge}$), 
gas density inside the edge ($n_{\rm in}$), power law index inside 
the edge ($\alpha_1$), and the IGM normalization ($K$) to vary. 
We find $r_{\rm edge} = 1.76$\,kpc, $\alpha_1 = -1.05$, and the 
discontinuity across the edge $J = 12.4$. 
This fit (Model A) 
is shown as the solid line in the top panel of 
Figure \ref{fig:n4783edgefit}.

The velocity of the galaxy is inferred from the pressure ratio 
$p_0/p_1$ between the undisturbed (free streaming) IGM and that 
just inside the cold front in the galaxy, that we use as a proxy 
for the IGM pressure  at the stagnation point. 
We use the
temperature of the IGM gas ($kT = 1.54^{+0.31}_{-0.22}$\,keV, 
abundance $0.46^{+0.19}_{-0.14}\Zs$), determined 
from the spectral model fit to angular sector $4$ in region P, 
(the region lying directly east of the leading edge of NGC~4783), as 
representative of the free streaming IGM in front of NGC~4783. 
We find a density discontinuity across the edge of 
$n_{\rm in}/n_{\rm out} = 10 \pm 3$
and a pressure ratio of $p_0/p_1 = 3.2^{+2.7}_{-1.8}$, 
where the uncertainties reflect both the uncertainties in 
temperature and abundance in our spectral fits.
We infer that NGC~4783 is moving at Mach 
$1.4^{+0.5}_{-0.7}$ relative to the IGM, or 
$v = 870^{+270}_{-400}$\kms relative to the group IGM. 
The upper limits  on the pressure ratio ($5.9$) and velocity
(Mach $1.9$, $v \sim 1140$\,kms) occur for the case of low abundance 
($0.3\Zs$), high temperature ($0.61$\,keV) galaxy gas in the galaxy's 
outer gas halo, coupled with low temperature ($\sim 1.32$\,keV) IGM
gas. The lower limit on the velocity (Mach$\sim 0.7$, 
$v \sim 470$\kms) requires high metallicity ($1.2\Zs$), low
temperature ($0.37$\,keV) galaxy gas. Physically, this latter
combination of parameters is unlikely to occur, since such high 
metallicities are most often present in the central cores of 
elliptical galaxies rather than at the contact discontinuity between 
the outer gas halo of the galaxy and the IGM. 

We note that the above hydrodynamic determination 
of the total velocity of NGC~4783 is independent of any possible
motion of the other interacting elliptical galaxy (NGC~4782) relative 
to the group gas. If, however, we assume that 
the measured relative radial velocity between NGC~4783 and NGC~4782 
($-623$\kms) is representative of the relative radial velocity of 
NGC~4783 with respect to the group gas (i.e. NGC~4782 is the 
dominant group elliptical galaxy and nearly at rest with 
respect to the group IGM), then we can constrain 
the three dimensional motion of NGC~4783 through the group. 
This assumption is supported by the dynamical measurements of group 
galaxies by Quintana \etal (1996) that place NGC~4782 at the systemic 
radial velocity of the group and is consistent with the X-ray surface 
brightness profile of the IGM gas, that is well fit by a spherically
symmetric $\beta$-model centered on NGC~4782 
(See \S\ref{sec:grouphydro} and Jones \etal 2007).
For a relative radial velocity $v_r = -623$\kms, our analysis implies a
transverse velocity for NGC~4783 of $600$ ($< 960$)\kms  
and an inclination angle  
$\zeta = 46^{\circ}$ ($> 33^{\circ}$) with respect to the plane of the
sky. In this model, the motion of NGC~4783 must be at least transonic 
(Mach $\geqslant 0.9$ for an IGM gas temperature
of  $1.54^{+0.31}_{-0.22}$\,keV), since the total velocity of the
galaxy must equal or exceed its radial velocity with respect to the IGM.    
   
This purely hydrodynamical determination of the 
velocity and inclination angle with respect to the group IGM 
(and NGC~4782) is in good agreement with the binary orbit parameters 
derived by Madejsky \& Bien (1993), who find a relative velocity between 
NGC~4783 and NGC~4782 of $\sim 1000$\kms and inclination angle for  
the plane of the orbit with respect to the plane of the sky of 
$\sim 40^\circ$ from numerical simulations of the tidal 
distortions of the stellar 
isophotes and velocities expected from the close encounter of two 
equal $\sim 5 \times 10^{11}\Ms$ galaxies. The binary orbital
parameters from Borne \etal (1988) favored a lower relative 
velocity ($\sim 700$\kms) and 
thus larger inclination angles ($\zeta \sim 65^\circ$). While our
analysis is also formally consistent with this set of orbital 
parameters, the sharpness of the observed surface brightness edge for 
NGC~4783, makes such a large inclination angle unlikely 
(Mazzotta \etal 2001). 

Our analysis for the region east of NGC~4783's leading edge  
is complicated by the presence of the radio lobe and associated 
X-ray cavity. We assumed that the
apparent coincidence of the leading edge of NGC~4783 with the outer 
contours of the radio lobe (see Figure \ref{fig:cavity}) is 
a projection effect, such that the pressure difference between the
galaxy gas inside the cold front and that in the free-streaming IGM 
is due solely to ram pressure caused by NGC~4783's motion through the 
IGM. We make a simple estimate of the maximal effect of the cavity 
on the surface brightness profile in front  of NGC~4783's edge by 
repeating the analysis using the X-ray decrement to rescale the 
observed surface brightness for radii $2 \lesssim r \lesssim 20$\,kpc 
(which overlap the X-ray cavity). 
This estimate for the maximal effect of the cavity on our analysis 
(Model B) reduces the inferred velocity of NGC~4783 and its upper 
uncertainty limit (in the steeply rising transonic/supersonic regime 
of Fig. \ref{fig:n4783edgefit}) by $\lesssim 10\%$, while the lower 
uncertainty limit is lowered by $\sim 24\%$.  

We also 
investigate whether NGC~4783 will ultimately merge with 
NGC~4782 by using the two body approximation to compute  the critical
mass $M_{\rm cr}$ 
required to lie within the radius of the orbit 
($r \sim 40''.6$) for NGC~4783 to be bound to NGC~4782. 
We find a critical mass of $(1.1 \pm 0.8) \times 10^{12}\Ms$ for
Model A, and lower bound of
$1.0^{+0.7}_{-0.8} \times 10^{12}\Ms$ for Model B, respectively.
Assuming that the group centered on NGC~4782 is in hydrostatic 
equilibrium and approximately isothermal, we can estimate  
the mass in the central region of the group using the $\beta$-model 
fit to the gas density  
for the group gas. Using the mean IGM temperature of $1.4$\,keV, 
we estimate  a total mass  of  $7 \times 10^{11}\Ms$ within the 
central $12.7$\,kpc ($40''.6$) of the group.  
If a gas temperature of
$2$\,keV is more representative of this central region, as suggested
by the lower right panel of Fig. \ref{fig:igm}, 
the total mass is $10^{12}\Ms$. 
After correcting for differences in the 
assumed cosmologies, these mass estimates are in good 
agreement with total mass estimates for NGC~4782 
($(7 \pm 2) \times 10^{11}\Ms$) obtained from the simulations of 
Borne \etal (1988), and also with the dynamical mass estimates 
($ 6 \times 10^{11}\Ms$ within $r \sim 3$\,kpc) of 
Madejsky \etal (1993). Within the uncertainties, these masses are
likely sufficient for NGC~4783 to be bound to NGC~4782.

\subsection{Kinematical Constraints from Bending the Eastern Radio Jet}
\label{sec:bend}

The morphology and evolution of radio jets offer 
another sensitive probe of galaxy and IGM motions.  
Assuming that ram pressure from the surrounding gas dominates, 
Euler's equation 
relates the ram pressure $\rho_{\rm ext} v_g^2$, where 
$\rho_{\rm ext}$ is the gas density of the surrounding gas and 
$v_g$ can be interpreted as either the relative velocity of the host 
galaxy through the ambient IGM or the velocity of local bulk
flows impacting the jets, to the velocity ($v_j$) and density 
($\rho_j$) of the radio plasma in the jet, the width of the 
jet ($h_j$) and the radius of curvature $R$ of the bend, i.e. 
\begin{equation}
\frac{\rho_j v_j^2}{R} = \frac{\rho_{\rm ext} v_g^2}{h_j}
\label{eq:rambend}
\end{equation}
(see, e.g., Eilek \etal 1984; Hardcastle \etal 2005a ). 
The measured minimum 
(equipartition) pressures of the radio plasma in the eastern and 
western jets are $1.1 \times 10^{-11}$ and $1.2 \times 10^{-11}$\ergcmc, 
respectively, and are a factor $\sim 3$ higher than the minimum 
pressure ($4.4 \times 10^{-12}$\ergcmc) measured in each of the radio
lobes. We use the minimum pressure in the eastern jet 
to estimate the minimum density $\rho_j = 3p_{\rm min}/c^2$ of plasma 
in the jet ($c$ is the speed of light), and determine the 
ram pressure velocity. 
Models for jet deceleration in FRI sources 
find that the flow within the jet is always transonic, with relativistic
Mach numbers inside the jet of $1.1-2$, corresponding to 
velocities of $(0.3 - 0.7)c$ for light, relativistic jets 
(Bicknell 1994; Laing \& Bridle 2002). 
 We adopt the mean velocity  
$(0.5 \pm 0.2)c$ as a reasonable estimate of the jet velocity in 
3C278 (NGC~4782), and interpret the observed lack of significant
beaming between the eastern and western jets as further evidence 
that the jets (and lobes) lie close to the plane of the sky. 
At the bend in the eastern jet, the jet width is $\sim 6''.4$ and the 
radius of curvature of the bend is $\sim 16''$. If the eastern 
jet is passing through the group IGM behind the `fan', we use the 
$\beta$-model fit to the group IGM around NGC~4782 
(from  \S\ref{sec:grouphydro} and Jones \etal 2007) to determine the  
external gas electron density at the location of the bend to be 
$7^{+0.8}_{-0.56} \times 10^{-3}$\cmc.
From equation \ref{eq:rambend},  
we find that a velocity of 
$v_g \sim 160 \pm 70$\kms 
would be sufficient to produce the observed
eastern jet bend. 

The second possibility is that the eastern jet passes through the 
disrupted ISM in the `fan region' (F). 
Using the lower bound on the density of the galaxy gas in the fan of  
$9^{+0.2}_{-0.3} \times 10^{-3}$\cmc 
(see Table \ref{tab:galdens.brems.2}), derived from 
our spectral models and  the assumption of uniform filling, 
for the external gas density acting on the jet in 
equation \ref{eq:rambend}, 
we find an upper limit on the velocity of  
$v_g < 140 ^{+100}_{-70}$\kms
for ram pressure bending of the jet. 
These velocities are modest, $\sim 20 - 60\%$ of the 
sound speed $c_s$ in the cool ($0.56$\,keV) galaxy gas in the fan. 
Thus, while they could be partly due to the motion of NGC~4782 
dragging the jet through the surrounding gas, they could also be due 
to bulk motions in the gas.  Madejsky (1992) found 
evidence of increased stellar velocity dispersion in NGC~4782 near the 
position of the jet bend, evidence that stellar mass 
in that region has been highly disturbed by the tidal 
encounter with NGC~4783. Tidal and pressure forces from that encounter
could have also induced bulk motions in 
the ISM gas at the location of the bend with comparable or 
greater velocities, i.e. $\sim 140$\kms, capable of bending the 
eastern jet. 

Since we do not have a direct measurement of the jet velocity and 
density of plasma inside the jet, it is useful to consider the 
behavior of our solutions as these parameters vary. 
In typical FRI sources initially light, highly relativistic 
($0.8-0.9c$) jets are thought to be decelerated to transonic 
velocities ($0.3-0.7c$) by the entrainment of external matter, with
density contrast $\rho_j/\rho_{\rm ext}$ growing from a few $\times 
10^{-6}$ to $10^{-4}$ by the end of the deceleration region 
(Bicknell 1994; Laing \etal 1999; Laing \& Bridle 2002). For the 
more powerful FRI sources, like 3C\,278 where the onset of flaring 
is $\gtrsim 2$\,kpc from the nucleus, the deceleration may extend 
well beyond $10$\,kpc, such that the density contrast at 
$5-6$\,kpc from the nucleus (at the bend in the eastern jet) is 
likely intermediate between these extremes. Assuming a density 
contrast at $5$\,kpc of $\sim 2 \times 10^{-5}$ in equation \ref{eq:rambend}, 
taken from the jet
deceleration model for 3C31 (see Fig. $7$ in Laing \& Bridle 2002), 
the ram pressure velocity needed to 
bend the eastern jet increases to $420 \pm 170$\kms, still subsonic 
or transonic with respect to the external medium and so not
unreasonable in a strongly interacting galaxy pair.
 
If the jet is much heavier and still fast, jet bending would require 
NGC~4782 to be moving supersonically with respect to the group IGM, 
which is not seen, or that highly supersonic fluid velocities impact the jets 
from the southeast as well as in the west, which is also unlikely. 
If the jets are light, but very non-relativistic (e.g. $\sim 0.03c$ as in 
Colina \& Borne 1993), then the velocities required to bend the 
jets would be very low ($\sim 10$\kms). Such jets would bend too 
easily and one would expect a narrow-angle tail (NAT) morphology 
rather than the observed mildly bent structure. Finally if the jet 
plasma is heavy, but slow, it would also have to be hot ($kT \gtrsim 20$\,keV)
such that the region containing the eastern lobe would still appear as an 
X-ray surface brightness decrement. However, if that were the case,
the jet would be far from equipartition, and the lobe highly  
overpressured with respect to the IGM. Thus it is most likely that the
jet plasma is light and relativistic, and is bent by subsonic 
bulk gas flows and/or motion of NGC~4782.
   
\subsection{Constraints from the Radio Lobes}
\label{sec:cavconst}

Properties of X-ray cavities (bubbles) associated with radio 
lobes place important constraints on the galaxy and/or gas
velocities in this system. For a cavity filled with relativistic 
plasma, the enthalpy of the cavity is $4pV$ where $p$ is the 
pressure of the hot gas evacuated from the cavity and $V$ is the 
cavity's volume. Using the IGM density distribution from 
\S\ref{sec:grouphydro} and assuming the cavity lies in the plane 
of the sky, we find an IGM electron density of $1.6 \times
10^{-3}$\cmc at $r=73''$($23$\,kpc), the radial distance from  
the nucleus of NGC~4782 to the center of the eastern lobe. For an 
IGM gas temperature of $1.54$\,keV (from the lower right panel 
of Fig. \ref{fig:igm}), 
the pressure of the hot gas  
evacuated from the expanding bubble was 
$\sim 7.5 \times 10^{-12}$\ergcmc. 
Note that this is less than a factor of two larger than the 
$4.4 \times 10^{-12}$\ergcmc measured minimum pressure inside the 
radio lobe. For a spherical volume of radius $34''$($10.7$\,kpc), 
the enthalpy carried in the eastern cavity is 
$\sim 4.4 \times 10^{57}$\,ergs.  If the cavity is inclined at 
$30^\circ$ with respect to the plane of the sky (the
maximum angle of inclination suggested by the X-ray decrement), the
density, pressure and bubble enthalpy each decrease by 
$\lesssim 10\%$.

We estimate the age of the cavity in two ways, using the geometry of the 
observed eastern cavity (see, e.g. Birzan \etal 2004; 
Dunn \etal 2005). First we obtain a lower bound on the age by assuming 
the bubble expands at the sound speed of the ambient medium ($\sim
640$\kms), which is likely most appropriate for young bubbles. 
Second, we use the time for the bubble to rise buoyantly, which is
likely appropriate when the cavity is more evolved.  We find a cavity 
age of $35 - 54$\,Myr, within a factor $1.4-3.7$ of the time since 
pericenter passage of NGC~4783 past NGC~4782 
(Borne \etal 1988; Madejsky \& Bien 1993). Thus  
the episode of AGN activity responsible for these features may 
have been influenced by the close encounter 
of the two galaxies. 
 Using these timescales and the enthalpy 
($4pV$) of the eastern cavity, we find a   
kinetic luminosity of $\sim 2.6 - 4 \times 10^{42}$\ergs. This is   
more than two orders of magnitude greater than the 
upper limit on the $0.5-10$\,keV X-ray luminosity for the AGN, 
estimated  from the spectral fits in \S\ref{sec:nucagn}, 
but still $\lesssim 4\%$ of the  
Bondi accretion power (assuming a $10\%$
efficiency) theoretically available to the AGN. 

Assuming no intrinsic sidedness, such that the enthalpy of the 
eastern and western lobes are equal, we can use the nonspherical 
geometry of the western lobe to constrain the external pressure 
exerted on the bubble. The projected geometry of the western 
lobe is an ellipse with semi-major (semi-minor) axes of $32''$ 
($18''$), respectively. Taking the three-dimensional 
geometry of the western cavity to be either an 
ellipsoid of rotation about the semi-major axis (prolate) or about the 
semi-minor axis (oblate), we can use the predicted enthalpy of the 
bubble ($4.4 \times 10^{57}$\ergs) to infer the external pressure 
against which the bubble had to expand. The pressures inferred from the
bubble enthalpy are 
$2.7 \times 10^{-11}$\ergcmc and $1.5 \times 10^{-11}$\ergcmc 
for the prolate and oblate geometries, 
respectively.  Using the IGM density 
distribution and mean temperature ($\sim 2.5$\,keV)
for gas to the west of NGC~4782 from \S\ref{sec:grouphydro},
 we measure a 
mean thermal pressure for IGM gas at the location of the western 
cavity of 
$1.4 \times 10^{-11}$\ergcmc.
The difference between the 
thermal pressure of the IGM and that predicted by the enthalpy 
carried in the bubble is interpreted as the ram pressure 
needed to distort the bubble geometry. For the prolate geometry this
implies a ram pressure velocity of 
$\sim 600 $\kms ($\sim 0.7c_s$ for 
$2.5$\,keV gas), a factor $\sim 3$ higher (depending on the assumed 
jet density), than the ram pressure 
velocity inferred from the eastern jet. 
However, the oblate (pancake) geometry may be more 
likely, since one might expect ram pressure from the southwest to 
cause the bubble to be compressed along the southern rim. For an 
oblate geometry, the predicted ram pressure velocity is lower, 
$\sim 170$\kms ($0.2c_s$ for $2.5$\,keV gas). Although the ram
pressure velocity, determined from the lobe analysis, 
 is likely uncertain by factors of a few,
due to the uncertainties in the lobe geometry and IGM gas
 temperatures, it is intriguing (and gratifying) that the ram pressure 
velocity for NGC~4782 required to bend the eastern jet 
($\sim 100 - 200$\kms) and western lobe ($\sim 170$\kms assuming
oblate geometry) are similar, and so likely result from the same 
dynamical processes.

\section{Discussion}
\label{sec:discuss}

The interaction picture of NGC~4782 and NGC~4783 that emerges from our 
analysis is the following:  
 The IGM gas temperature and X-ray surface
 brightness distribution show that NGC~4782 and NGC~4783 reside in 
a large, massive galaxy group (LGG~316) with elliptical galaxy 
NGC~4782 likely the dominant group galaxy near the center of LGG~316's 
deep, dark matter potential. The lack of a sharp X-ray surface
brightness edge to the south of NGC~4782 supports the interpretation 
that NGC~4782 is the dominant group galaxy and is not moving rapidly
with respect to the center of the group. In contrast, analysis of the 
sharp X-ray surface brightness edge on the eastern (leading) side of
the elliptical galaxy NGC~4783, coupled with its narrow, trailing 
 ram-pressure stripped tail, shows NGC~4783 infalling supersonically 
 through the group IGM. As NGC~4783 passes supersonically from 
 southwest to east past NGC~4782, tidal interactions distort the 
 spatial and velocity distributions of stars and gas in and between 
 the galaxies.  In addition, the compression wave produced in the 
 IGM by NGC~4783's 
 high speed passage through the group gas impacts NGC~4782's ISM, 
 displacing it to the east relative to the stellar
 isophotes and creating the `fan'.  These bulk gas velocities, perhaps 
 aided by a small tidally-induced 
 oscillation of NGC~4782 relative to the center of the group's 
 gravitational potential, provide the ram-pressure needed to bend 
 the radio jets. Our analysis demonstrates that understanding 
 the hydrodynamical interactions between dumbbell galaxies and 
 their surrounding gas environment, as revealed by X-ray observations, 
 plays a critical role in constraining the interaction kinematics of 
 these systems and in modeling their evolution in groups, as well as
 in richer cluster environments. 

\section{Conclusion}
\label{sec:conclude}

In this paper we have used the results of a $49.3$\,ks {\it Chandra} X-ray 
observation of the dumbbell galaxies NGC~4782 and
NGC~4783 to measure the thermodynamic properties of hot gas in and 
surrounding the interacting system and have compared our results with 
new $1.5$ and $4.9$\,GHz VLA maps of the dumbbell. We then used these 
properties to constrain the kinematics of the interaction and  models for the
bending of the radio jets associated with host galaxy NGC~4782. 
We find the following:

\begin{itemize}

\item{In the interaction region close to the two galaxies, the
X-ray emission is highly asymmetric, showing features characteristic 
of both tidal and gas-dynamical stripping. NGC~4783 possesses a sharp 
leading surface-brightness edge $1.76$\,kpc east of its
center and a $\sim 15$\,kpc tail extending to 
the west, characteristic of ram pressure stripping, and fixing the 
direction of motion of NGC~4783 in the plane of the sky to be 
from west to east.}

\item{The temperature, density, and pressure ratios between gas 
inside NGC~4783's leading edge and undisturbed 
IGM gas in front of NGC~4783 
indicate that NGC~4783 is moving at velocity 
$\sim 870^{+270}_{-400}$\kms (Mach $1.4^{+0.5}_{-0.7}$) with
respect to the group gas. Assuming a radial velocity of $-623$\kms  
( NGC~4782 radially at rest with respect to the IGM), we find an 
inclination angle for NGC~4783's motion of $46^\circ (> 33^\circ)$  
with respect to the plane of the sky. This is in agreement with 
the binary orbital parameters of Madejsky \& Bien (1993) found by 
fitting tidal distortions in the stellar isophotes and velocities. 
}
 
\item{ A tidal bridge of X-ray emission extends along the line 
between the centers of NGC~4782 and NGC~4783, and a fan of 
X-ray emission extends to the east 
of NGC~4782. Hot gas in the central regions of both galaxies is well 
fit by single 
temperature APEC models with $kT \sim 0.4-0.5$\,keV, while the tail and 
bridge temperatures are higher ($kT \sim 0.68$\,keV). 
The `fan' to the east of NGC~4782 is composed of
cool ($0.56$\,keV) galaxy gas. 
The brightest X-ray emission in the fan is near the base of the
eastern radio lobe.
}

\item{ Outside the interaction region, the X-ray surface brightness 
 of the group IGM is approximately spherically symmetric and nearly 
 isothermal with mean temperature $\sim 1.4 \pm 0.4$\,keV. The IGM 
 surface brightness is well fit by a power law density distribution 
 $n_e = 2.5^{+0.3}_{-0.2} \times 10^{-3}(r/15.45\,{\rm kpc})^{-3\beta}$\cmc 
  with $\beta = 0.35$, 
 centered on NGC~4782.  However, the data suggest that close to the 
 interaction region, at about the radii of the radio lobes 
 ($17 \lesssim r \lesssim 37$\,kpc),
 the mean gas temperature to the west of the dumbbell is higher 
 ($2.51^{+0.55}_{-0.38}$\,keV) than the gas temperature to the east
 ($1.75^{+0.25}_{-0.13}$\,keV). We find  the group X-ray 
 luminosity  and total mass within a radius of $100$\,kpc of
 NGC~4782's center to be 
  ${\rm log}\,L_{\rm X} = 41.99 ^{+0.13}_{-0.04}$)\ergs
 and $(5.5 \pm 1.5) \times 10^{12}\Ms$, respectively.
 Using the M-T relation from Finoguenov \etal (2001), we 
 estimate a total mass (at $r_{500}$) for the group of $4.8^{+2.7}_{-2.2} 
 \times 10^{13}\Ms$, consistent with the large ($447$\kms) galaxy
 velocity dispersion measured for the group as a whole. X-ray 
 morphology and gas temperatures at $r > 37$\,kpc 
 favor the interaction picture in which the elliptical galaxy NGC~4783 
 is infalling for the first time into a single massive group  
 with NGC~4782 nearly at rest at the center of the group potential.}

\item{There is a $29^{+12}_{-13}\,\%$ ($1\sigma$ uncertainties) 
 reduction in the X-ray surface brightness 
 in the region coincident with the eastern radio lobe of NGC~4782, 
 indicating the presence of an X-ray cavity. The observed X-ray 
 surface brightness decrement suggests that the inclination angle of the 
 eastern lobe is $\lesssim 30^\circ$ with respect to the plane of the 
 sky. Assuming relativistic plasma fills the lobe, the enthalpy
 carried in the cavity is $\sim 4.4 \times 10^{57}$\,ergs. The 
 age of the eastern cavity is $\sim 35 - 54$\,Myr, and its 
 kinetic power  is $\sim 2.6 - 4 \times 10^{42}$\ergs.
 }

\item{The eastern radio jet of 3C278 bends sharply to the northeast 
at $\sim 5-6$\,kpc from 
the nucleus of NGC~4782, inflating shortly thereafter into a roughly 
spherical lobe of radius $\sim 10.7$\,kpc, coincident with the X-ray 
cavity. The western jet extends straight $\sim 12.6$\,kpc before 
flaring into an ellipsoidal lobe that bends gradually to the 
northwest. The minimum radio pressures in the eastern and western 
jets are $1.1 \times 10^{-11}$ and $1.2 \times 10^{-11}$\ergcmc, 
respectively, and $4.4 \times 10^{-12}$\ergcmc in each of the lobes.
These equipartition pressures are within factors $\sim 2 - 3$ of the 
pressures of the surrounding hot IGM gas.}

\item{X-ray knots are  found at the end of the eastern inner radio jet
and in the western inner radio jet, coincident with 
peaks in the $4.9$\,GHz radio emission, and are likely 
associated with deceleration of the radio plasma. }

\item{A ULX with $L_{\rm X}  \sim  3 \times 10^{39}$\ergs is found 
near NGC~4782's nucleus. } 
  
\item{ Assuming a light, mildly relativistic jet,  
ram pressure velocities of $\sim 100 - 200$\kms are sufficient 
to produce the observed bending in the eastern jet. Assuming no 
intrinsic jet sidedness and an oblate 3-dimensional geometry
for the western lobe, we find the ram pressure velocity of the 
IGM relative to the western lobe to be $\sim 170$\kms with
a factor of order $\sim 2$ uncertainty. These velocities may reflect  
bulk motions established in the gas during the galaxies' high velocity  
encounter, with a possible additional small contribution caused by the  
oscillation of NGC~4782 with respect to the group potential, also 
induced by the tidal interaction of the galaxy pair.} 

\end{itemize}

\acknowledgements

This work is supported in part by NASA grant AR5-6011X, 
the Smithsonian Institution and the Royal Society.
The National Radio Astronomy Observatory is a facility of the
National Science Foundation operated under cooperative agreement by
Associated Universities, Inc. This work has made use of the NASA/IPAC 
Extragalactic Database (NED)
which is operated by the Jet Propulsion Laboratory, California
Institute of Technology,  under contract with the National
Aeronautics and Space Administration. We wish to thank Karen Masters
for help using the CfA redshift survery, John Huchra and Nathalie 
Martimbeau for new redshift measurements of NGC~4782 and  NGC~4783, 
and Maxim Markevitch for helpful discussions and 
use of his edge fitting analysis codes.
 

\clearpage

\begin{deluxetable}{llllr}
\tablecaption{VLA observations used in this paper}
\tablehead{Program ID&VLA configuration&Frequency (GHz)&Date&Time on source
(min)}
\startdata
AC104&C&1.46, 1.51&1984 May 03&11\\
AC104&C&4.84, 4.89&1984 May 03&147\\
AC104&A&4.84, 4.89&1984 Dec 26&211\\
AC104&B&1.46, 1.51&1985 May 13&43\\
AC104&B&4.84, 4.89&1985 May 13&221\\
AH343&D&4.84, 4.89&1989 Nov 17&9\\
\enddata
\label{vla-data}
\tablecomments{All observations were made at two observing
  frequencies, quoted in this table. The effective frequency, the
  mean of these two frequencies, is used in the text.}
\end{deluxetable}

\begin{deluxetable}{lcccc}
\tablewidth{0pc}
\tablecaption{Spectral Analysis Regions for NGC~4782 and 
 NGC~4783\label{tab:specregsv2}}
\tablehead{
\colhead{Region } &\colhead{Shape} &\colhead{Center}&
\colhead{Dimensions} &\colhead{Orientation}  \\
  &   & R.A. Dec. &      &   \\
  &   & (J2000.0) & (arcsec) & (deg)    }
\startdata
H  & circular    &$12\,\,54\,\,36.6$  $-12\,\,33\,\,28.2$ & $7$& \ldots \\
H$_{\rm bg}$ &  &  & & \\
   & circular    &$12\,\,54\,\,37.6$  $-12\,\,34\,\,36.8$ & $7$& \ldots \\
   & circular    &$12\,\,54\,\,33.9$  $-12\,\,34\,\,36.8$ & $7$& \ldots \\
   & circular    &$12\,\,54\,\,36.3$  $-12\,\,34\,\,46.2$ & $7$& \ldots \\
   & circular    &$12\,\,54\,\,35.1$  $-12\,\,34\,\,46.2$ & $7$& \ldots \\
T & rectangular &$12\,\,54\,\,34.8$  $-12\,\,33\,\,36.3$ & $36$,$17$ &$333$ \\
T$_{\rm bg}$& rectangular &$12\,\,54\,\,36.6$  $-12\,\,34\,\,37.3$ & $36$,$17$ &$333$ \\
B  & rectangular &$12\,\,54\,\,36.4$ $-12\,\,33\,\,49.3$ & $20$, $14$ &$113$\\
B$_{\rm bg}$  & rectangular &$12\,\,54\,\,35.1$ $-12\,\,34\,\,24.5$ & $20$, $14$ &$113$\\
F  & rectangular &$12\,\,54\,\,37.3$  $-12\,\,33\,\,59.4$ & $25$, $18$ &$294$ \\
F$_{\rm bg}$ & rectangular &$12\,\,54\,\,36.5$ $-12\,\,34\,\,29.2$& $25$, $18$ &$294$ \\
N  & circular    &$12\,\,54\,\,35.7$  $-12\,\,34\,\,06.6$ & $2.5$ & \ldots \\
N$_{\rm bg}$  & annular &$12\,\,54\,\,35.7$  $-12\,\,34\,\,06.6$ & $3$, $11$ & \ldots\\
A  & annular     &$12\,\,54\,\,35.7$  $-12\,\,34\,\,06.6$ & $3$, $11$ & \ldots\\
A$_{\rm bg}$ &annular& $12\,\,54\,\,35.7$ $-12\,\,34\,\,06.6$ & $12$, $25$ & $195 - 340$ \\
\enddata
\tablecomments{ Dimensions specified are radii for 
  circular regions, (inner, outer) radii for annular regions, and 
(length,width) for rectangular regions.  Local background regions are
  specified by the subscript bg. Orientation angles are measured
 counterclockwise from west for the rectangular region's major (length)
axis and the annular sector A$_{\rm bg}$ (see Fig. \protect\ref{fig:galspecv2}).
}
\end{deluxetable}

\begin{deluxetable}{cccc}
\tablewidth{0pc}
\tablecaption{Spectral Fits for Elliptical Galaxies NGC~4782 and NGC~4783 
\label{tab:galspectra.brems.2}}
\tablehead{\colhead{Region} &
\colhead{Source} &\colhead{$kT_1$} &  
  \colhead{$\chi^2/{\rm dof}$} \\
  & counts & keV & }
\startdata
H   &$195  \pm 15$ & $0.48^{+0.13}_{-0.11}$ &$5.8/8$ \\
T   &$290 \pm 26$  & $0.68^{+0.05}_{-0.07}$ &$23.6/19$  \\
B  &$208 \pm 19$  & $0.68^{+0.07}_{-0.08}$ & $13.3/10$ \\
F  &$146 \pm 20$  & $0.56^{+0.13 }_{-0.14}$ & $2.7/10$ \\
A  &$213 \pm 20 $  & $0.43^{+0.16}_{-0.10}$ & $10.9/13 $ \\
\enddata
\tablecomments{  Col. $2$ lists the net source counts in the 
$0.3 - 8$\,keV range for regions H, T, A and $0.3-2$\,keV for 
regions B and F.  
The spectral models are absorbed APEC models 
with an additional fixed $5$\,keV bremsstrahlung component to model 
the contribution from unresolved LMXBs.
The hydrogen absorbing column is fixed at the Galactic value 
($n_{\rm H}=3.58 \times 10^{20}$\cms; Dickey \& Lockman 1990) for all
model components in each region and metal abundances for the APEC 
thermal plasma components are fixed at $A = 0.5\,\Zs$. The
temperatures are unchanged when the abundance is varied between
$0.3-1.2\Zs$. Errors correspond to $90\%$ confidence limits. 
}
\end{deluxetable}

\begin{deluxetable}{ccccc}
\tablewidth{0pc}
\tablecaption{X-ray Luminosities for Regions in 
 NGC~4782/NGC~4783\label{tab:lums}}
\tablehead{\colhead{Region} &\colhead{$L_{\rm gas}$}& \colhead{$L_{\rm gas}$}
 &\colhead{$L_{\rm LMXB}$}& \colhead{$L_{\rm LMXB}$}\\
 & ($0.5-2$\,keV)&($2-10$\,keV) & ($0.5-2$\,keV) & ($2-10$\,keV)}
\startdata
H &$5.4$ &$0.1$ &$2.5$ &$3.8$ \\
T &$9.4$ &$0.3$ &$1.3$ &$2.0$\\
B &$7.3$ &$0.3$ &$1.4$ &$2.1$ \\
F &$5.0$ &$0.1$ &$0.9$ & $1.4$ \\
A &$3.4$ &$0.03$ &$4.7$ &$6.9$ \\
\enddata
\tablecomments{Intrinsic X-ray luminosities in units of $10^{39}$\ergs
  for the galaxy emission regions shown in Fig. \protect\ref{fig:galspecv2} 
 and spectral models listed in Table \protect\ref{tab:galspectra.brems.2} and 
assumed luminosity distance of $66.7$\,Mpc. The quoted values for the 
$0.5-2$ ($2-10$)\,keV luminosities, that assume gas metallicity
$0.5\Zs$, change by $\lesssim 4\%$($17\%$) when the abundance is 
varied between $0.3 - 1.2\Zs$.
} 
\end{deluxetable}

\begin{deluxetable}{ccccc}
\tablewidth{0pc}
\tablecaption{Electron Densities, Thermal Pressures and Gas Masses  
for Regions in NGC~4782/NGC~4783\label{tab:galdens.brems.2}}
\tablehead{\colhead{Region} &\colhead{$n_e$} &\colhead{$p$} 
&\colhead{$M_g$} &\colhead{$t_c$}\\
  & ($10^{-2}$\cmc) & ($10^{-11}$ergs\,\cmc) & ($10^7\,\Ms$) &(Gyr) \\
(1)&(2)&(3)&(4)& (5)}
\startdata
H &$2.2^{+0.5}_{-0.8}$ &$3.3^{+1.8}_{-1.6}$ &$2.9^{+0.6}_{-1.0}$ &$0.34^{+0.08}_{-0.11}$ \\
T  &$1.1^{+0.2}_{-0.4}$& $2.2^{+0.6}_{-0.9}$&$8.0^{+1.7}_{-2.6}$ & $0.86^{+0.26}_{-0.33}$ \\ 
B  &$1.6^{+0.3}_{-0.5}$ &$3.3^{+1.1}_{-1.3}$ &$4.2^{+0.9}_{-1.4}$ &$0.57^{+0.19}_{-0.22}$ \\  
F &$0.9^{+0.2}_{-0.3}$ &$1.5^{+0.9}_{-0.7}$ &$5.3^{+1.2}_{-1.8}$ &$0.89^{+0.46}_{-0.44}$ \\
A  &$0.9 \pm 0.3$ &$1.3^{+0.9}_{-0.7}$ &$4.4^{+1.1}_{-1.5}$ & $0.77^{+0.54}_{-0.37}$\\
\enddata
\tablecomments{Col.(1): region identifier, col.(2): electron density,
  col. (3): thermal gas pressure, col. (4): gas mass, col. (5):
  gas cooling time.  Spectral models are given in Table
  \protect\ref{tab:galspectra.brems.2} for the spectral regions 
   defined in Table \protect\ref{tab:specregsv2}. Values are listed  
   for uniform filling and zero inclination angle with respect 
   to the plane of the sky. Uncertainties reflect both the $90\%$ CL 
   uncertainty in the gas temperature and the uncertainty from varying 
   the gas metal abundance between $0.3-1.2\Zs$ in the spectral models.
} 
 \end{deluxetable}

\begin{deluxetable}{ccccc}
\tablewidth{0pc}
\tablecaption{Nuclear Region Spectral Fits\label{tab:nucspecfit}}
\tablehead{\colhead{Region} &
\colhead{Source} &\colhead{$kT_1$} &\colhead{$\Gamma$}&  
  \colhead{$\chi^2/{\rm dof}$} \\
  & counts & keV & & }
\startdata
N &$ 193 \pm 14.5$ & $0.58$ & \ldots  &$26/8$ \\
N &$ 193 \pm 14.5$ & \ldots & $3.44$ &$21/8$ \\
N &$ 193 \pm 14.5$ & $0.43^{+0.17}_{-0.12}$ & $1.7f$ &$7.0/7$ \\
\enddata
\tablecomments{Col. $2$ lists the net source counts in the $0.3-5$\,keV 
energy band for region N.  The spectral models are absorbed 
APEC, powerlaw, or APEC + powerlaw 
with an additional $5$\,keV bremsstrahlung component to model the
contribution from unresolved 
LMXBs.  The hydrogen absorbing column is fixed at the Galactic value 
($n_{\rm H}=3.58 \times 10^{20}$\cms; Dickey \& Lockman 1990) for all
model components in each region and metal abundances for the APEC 
thermal plasma components are fixed at $A = 0.5\,\Zs$.
Errors correspond to  $90\%$ confidence limits. $f$ denotes a fixed 
parameter.}
\end{deluxetable}

\begin{deluxetable}{cccc}
\tablewidth{0pc}
\tablecaption{Group Properties\label{tab:grpprop}}
\tablehead{\colhead{Row} & \colhead{Property} & \colhead{Value} & \colhead{Units}}
\startdata
(1) &$\sigma_v$ & $447 \pm 70$ & \kms \\
(2) &$r_{500} $ & $0.66$& Mpc \\
(3) &$M_{\rm dyn}(r_{500})$ & $7 \times 10^{13}$ &$\Ms$ \\
(4) &$M_{\rm scale}(r_{500}$ & ($4.8^{+2.7}_{-2.2}) \times 10^{13}$ & $\Ms$ \\
(5) &$\beta$ & $0.35$ & \\
(6) &$n_{\rm IGM}$ & $(2.5^{+0.3}_{-0.2}) \times 10^{-3}$& \cmc \\
(7) &$kT_{{\rm mean}}$ & $1.4 \pm 0.4$ & keV \\
(8) &$A$ & $0.3-0.5$ & $\Zs$ \\
(9) &${\rm log}\,L_{\rm X}(100\,{\rm kpc})$ &
$41.99^{+0.13}_{-0.14}$& \ergs \\
(10) &${\rm log}\,L_{\rm X}(r_{500})$ & $42.73^{+0.13}_{-0.05}$& \ergs \\
\enddata
\tablecomments{Row (1) LOS velocity dispersion, row (2) $r_{500}$ 
 from Eq. 1, Osmond \& Ponman (2004), row (3) dynamical mass, row(4) mass  
 from $M-T$ scaling relation for groups (Finoquenov \etal (2001)), 
 row (5) and (6) IGM density model parameters ($n_e = n_{\rm
  IGM}(r/15.45\,{\rm kpc})^{-3\beta}$), row (7) and (8) IGM mean 
 temperature and metal abundance, row (9) measured total X-ray luminosity 
 within $r \leq 100$\,kpc, row (10) extrapolated total  
  X-ray luminosity within $r_{500}$.}  
\end{deluxetable}

\begin{deluxetable}{cccc}
\tablewidth{0pc}
\tablecaption{IGM Region Geometry\label{tab:pandageom}}
\tablehead{\colhead{Annulus} & \colhead{Radii} & 
\colhead{Angular Extent} & \colhead{Annular Sectors} \\
  &r$_{in}$, r$_{out}$&
      $\theta_{\rm start}$, $\theta_{\rm stop}$& \\
  (1)& (2) & (3) & (4) }
\startdata
P$_s$ & $15$, $54$ & $185$, $25$ & $6-9,1$ \\
P & $54$,$118$ & $345$, $345$ & $1 - 9$ \\
P$_{n1}$& $118$, $187$ & $345$, $225$ & $1 - 6$ \\
P$_{n2}$ &$187$, $236$ & $65$, $145$ & $3,4$ \\
P$_{n3}$ &$236$, $305$ & $65$, $145$ & $3,4$ \\
\enddata
\tablecomments{Col. (1) annulus region identifier, col. (2) inner
  radius, outer radius of the annulus in arcsec, col. (3)
  beginning angle, ending angle for the  annulus in degrees 
  measured counterclockwise from west, col. (4) $40^\circ$ annular  
  sectors numbered counter clockwise from $345^\circ$ (see Fig.
  \protect\ref{fig:igm}.
  The annuli are concentric with 
   centers at $12^h54^m35.8^s$,$-12^\circ34'7.3''$ (J2000.0)
     } 
\end{deluxetable}

\begin{deluxetable}{ccccc}
\tablewidth{0pc}
\tablecaption{Gas Temperatures and Abundances in the IGM\label{tab:igmfitswO}}
\tablehead{\colhead{Region} &
\colhead{Source} &\colhead{$kT$} &\colhead{$A$} & 
  \colhead{$\chi^2/{\rm dof}$} \\
  & counts & keV & $\Zs$ &  }
\startdata
P$_s$  &$911 \pm 36$  &$1.62^{+0.31}_{-0.13}$ &$0.55^{+0.3}_{-0.16}$ &$47.5/49$
 \\
P    &$2870 \pm 73$  &$1.98^{+0.17}_{-0.22}$ &$0.46^{+0.19}_{-0.14}$ &$176.7
/171$ \\
P$_{n1}$ &$2269 \pm 67$ &$1.87 \pm 0.28$ &$0.28^{+0.22}_{-0.15}$ &$121/133$ \\
P$_{n2}$ &$416 \pm 38$ &$1.21^{+0.23}_{-0.20}$ &$0.31^{+0.72}_{-0.18}$ &$40.3/58
$ \\
P$_{n3}$ &$766 \pm 52$ &$1.37^{+0.38}_{-0.22}$ &$0.15^{+0.26}_{-0.09}$ &$115/100
$ \\
\enddata
\tablecomments{ Col. (2) lists the net source counts in the 
$0.3-8$\,keV energy band. All regions used an absorbed APEC spectral
  model for the IGM with hydrogen absorption fixed at the Galactic value, 
$n_{\rm H} = 3.58 \times 10^{20}$\cmc.
 }
\end{deluxetable}

\end{document}